\documentclass[12pt,onecolumn,draftcls]{IEEEtran}
\usepackage{graphicx}
\usepackage{bm}
\usepackage[cmex10]{amsmath}
\usepackage{amssymb}
\usepackage{acronym}
\usepackage{setspace}
\usepackage{amsthm}
\usepackage{algorithm}  
\usepackage{algorithmic} 
\usepackage{footnote}
\usepackage{cite}

\newtheorem{theorem}{Theorem}

\ifCLASSINFOpdf
\else
\fi

\begin{document}
\title{MIMO Multiway Distributed-Relay Channel with Full Data Exchange: An Achievable Rate Perspective}
\author{
	Xiang~Zhao,
    Jianwen~Zhang,
    Ying Jun (Angela) Zhang,~\IEEEmembership{Senior Member,~IEEE},
    Xiaojun~Yuan,~\IEEEmembership{Senior Member,~IEEE}
\thanks{X. Zhao, J. Zhang and X. Yuan are with the School of Information Science and Technology, ShanghaiTech University, Shanghai, China, email: \{zhaoxiang, zhangjw1, yuanxj\}@shanghaitech.edu.cn.}
\thanks{Y. Zhang is with the the Department of Information Engineering, The Chinese University of Hong Kong, Shatin, New Territories, Hong Kong, e-mail: yjzhang@ie.cuhk.edu.hk.}
\thanks{The work was partially presented in IEEE International Conference on Communications in China (ICCC2016).}
}


\maketitle

\begin{abstract}
		We consider efficient communications over the multiple-input multiple-output (MIMO) multiway distributed relay channel (MDRC) with full data exchange, where each user, equipped with multiple antennas, broadcasts messages to all the other users via the help of a number of distributive relays. We propose a physical-layer network coding (PNC) based scheme involving linear precoding for channel alignment, nested lattice coding for PNC, and lattice-based precoding for interference mitigation. We show that, with the proposed scheme, distributed relaying achieves the same sum-rate as cooperative relaying in the high SNR regime. We also show that the proposed scheme achieves the asymptotic sum capacity of the MIMO MDRC within a constant gap in the high SNR regime. Numerical results demonstrate that the proposed scheme considerably outperforms the existing schemes including decode-and-forward and amplify-and-forward.
\end{abstract}


\IEEEpeerreviewmaketitle

\section{Introduction}
	\IEEEPARstart{P}{hysical}-layer network coding (PNC) has proved a great success in enhancing the throughput of wireless relay networks in the past decade \cite{Zhang06,Yuanxj2012,NamIT10,YangIT11,YangTCOM13}. An early application of PNC is the so-called two-way relaying \cite{Zhang06}\cite{Yuanxj2012}, where two users exchange information via the help of a single relay. It was shown in \cite{NamIT10} that PNC with nested lattice coding \cite{NamIT10} can achieve the capacity of the two-way relay channel (TWRC) within $\frac{1}{2}$ bit. Later, multiple-input multiple-output (MIMO) techniques were introduced into the TWRC to achieve multiplexing gain. It has been revealed in \cite{YangIT11,YangTCOM13,YuanIT13,KhinaISIT11} that near-capacity performance can be realized by using a sophisticated combination of advanced signal processing and coding techniques including linear precoding, nested lattice coding, and successive interference cancellation, etc.	
		
	As a natural extension, much research interest has been attracted to a more general relay model, termed \textit{MIMO multiway relay channel} (MRC), in which a number of multi-antenna users exchange information via the help of a multi-antenna relay node \cite{mul-communication,multiway,Wang16,Tian14}. This setup models many practical communication scenarios, e.g., in an ad hoc network, a group of users want to share files with the help of a relay while each user only has a distinct portion of a common file desired by all the other users. Various data exchange models of the MIMO MRC have been investigated in the literature \cite{other.data.exchange.model.1,Xin13,other.data.exchange.model.2}. In particular, the authors in \cite{LeeIT10} and \cite{Wang15'1} considered pairwise data exchange in which users exchange messages in a pairwise manner; the works in \cite{Yuanxj2} and \cite{Huang14} assumed a full data exchange model in which each user wants to learn the messages from all the other users. Other variants, such as X channels and Y channels, have also been studied in \cite{DingPre,LeeTWC12,Liu}. In these works, advanced signal alignment techniques were developed to jointly design the user precoders and the relay precoder for efficient PNC implementation. Based on that, the degrees of freedom (DoF) is analyzed for the MIMO MRC with different network topological configurations and antenna settings.
	
	The DoF analysis characterizes the system performance in the high signal-to-noise ratio (SNR) regime. This characterization is insufficient from a practical point of view, since practical communication systems usually work at low and medium SNR. As such, many research groups have attempted to analyze the fundamental capacity limits of the MIMO MRC. This is a challenging task since the capacity characterization of even the simplest three-node relay model \cite{Cover} has remained an open problem for decades. In this research line, some initial capacity results of the MIMO MRC were reported in \cite{cyclic,Ong11,Fang14}. For instance, the authors in \cite{Fang14} showed that the asymptotic capacity of the cellular MIMO TWRC is achievable by using linear precoding and nested-lattice coding techniques.
	
	The works mentioned so far are limited to the configuration of a single relay. In these networks, the relay node is usually the performance bottleneck since all the traffic flows need to go through the relay. Distributed relaying breaks this limitation, so as to boost the network throughput \cite{Wang15'2,dis.antenna.1,dis.antenna.2,dis.antenna.3}. In particular, the authors in \cite{LeeTWC12} studied the DoF of the MIMO multipair TWRC with multiple relays, where a general framework for the DoF analysis was established by combining the ideas of signal space alignment and interference neutralization. However, as aforementioned, the DoF characterization is usually not sufficient for understanding the behavior of the network in the practical SNR regime. As such, the capacity analysis of distributed multiway relaying is highly desirable.
	
	In this paper, we investigate the design of efficient communication strategies to approach the capacity of the MIMO multiway distributed-relay channel (MDRC) with full data exchange. The proposed scheme involves linear precoding for channel alignment, nested lattice coding for PNC, and lattice-based precoding for interference mitigation. We derive an achievable rate region of the proposed scheme. We show that, as compared with cooperative relaying, distributed relaying (in which relays do not share their received signals) does not suffer any rate loss in the high SNR regime. We also show that our proposed scheme achieves the asymptotic capacity of the MIMO MDRC within a constant gap at high SNR. In particular, this gap vanishes in the two-user case, i.e., our proposed scheme can achieve the asymptotic capacity of the MIMO TWRC with distributed relays. Numerical results demonstrate that the proposed scheme considerably outperforms the existing relaying schemes including decode-and-forward \cite{df.1},\cite{df.2} and amplify-and-forward \cite{dis.antenna.2},\cite{af.1}.
	
	It is worth mentioning that the asymptotic capacity of the MIMO TWRC was previously achieved by using the generalized singular value decomposition (GSVD) based nested lattice coding scheme proposed in \cite{YangIT11}. However, GSVD requires relay cooperation and hence cannot be applied to the case of distributed relays. Therefore, the result in this paper is the first to achieve the asymptotic capacity of the MIMO TWRC with distributed relays.
	
	The remainder of this paper is organized as follows. In Section \ref{sec model}, we describe the system model. In Section \ref{sec proposed}, we introduce our proposed relay protocol including the operations at both users and relays, and derive an achievable rate region for the overall system. In Section \ref{sec sumrate}, we consider the sum-rate maximization problem. The asymptotic analysis is presented in Section \ref{sec asy} and the optimality of the proposed scheme is also given in this section. The numerical results are presented in Section \ref{sec num}. Finally, we conclude the paper in Section \ref{sec con}.
	
	Regular letters, lowercase bold letters, and capital bold letters represent scalars, vectors, and matrices, respectively. For any matrix $\mathbf{A}$, let $ \mathbf{A}^{\text{T}} $, $\text{tr}\{\mathbf{A}\}$, and $|\mathbf{A}|$ be the transpose, the trace, and the determinant of $\mathbf{A}$, respectively. $\|\cdot\|_2$ represents the $\ell_2$-norm; $\text{E}[\cdot]$ denotes the expectation operation; $\log (\cdot )$ denotes the logarithm with base 2; $[\cdot ]^{+}$ denotes $\max \{\cdot ,0\}$. For any matrices $\mathbf{A}$ and $\mathbf{B}$, $\mathbf{A}\succeq \mathbf{B}$ means $\mathbf{A} - \mathbf{B}$ is semi-positive definite. For an integer $N$, $\mathcal{I}_N$ denotes the set of integers from $ 1 $ to $ N $; $\mathcal{\mathbb{R}}^{n\times m}$ denotes the $n$-by-$m$ dimensional real space;  $\mathcal{N}(\mu ,\sigma ^{2})$ denotes the Gaussian distribution with mean $\mu$ and variance  $\sigma^{2}$.


\section{System Model}\label{sec model}

\subsection{System Model}
\begin{figure}[tp]
	\centering
	\includegraphics[width=0.6\textwidth]{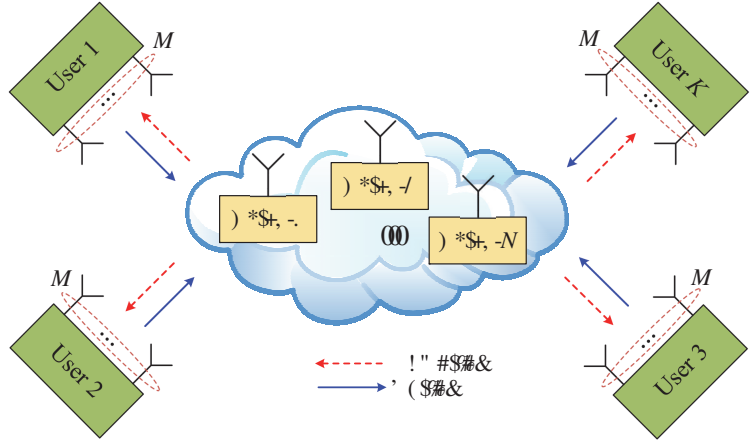}
	\caption{The system model of a MIMO MDRC channel with $K$ users and $N$ relays. Each user is equipped with $M$ antennas.}
	\label{Figure 1}
\end{figure}

Consider a discrete memoryless MIMO multiway distributed-relay channel (MDRC), where $K$ users, each equipped with $M$ antennas, broadcast their messages to all the other users with the help of $N$ single-antenna relays, as illustrated in Fig. \ref{Figure 1}. The data exchange pattern is assumed to be full data exchange, i.e., each user desires messages from all other $K-1$ users\cite{Huang14},\cite{Ong11}. We assume that there is no direct link between any two users. This assumption can be justified by practical communication scenarios with severe path attenuation and line-of-sight obstruction. We also assume that the network operates in a half-duplex mode, i.e., a node cannot transmit and receive signal simultaneously at a common frequency band.
Each round of data exchange  consists of two phases, termed the uplink phase and the downlink phase. Without loss of generality, we assume that the time duration of the uplink phase is $ \alpha T $ and that of the downlink is $ (1-\alpha)T $, where $ \alpha  \in (0,1)$ is a time-splitting  factor, and $ T $ is the time duration of each round of data exchange.  In the uplink phase, all $K$ users simultaneously transmit signals to the relays. The received signal of relay $n$ at time $t$ is given by 
\begin{equation}\label{eq.yRn}
	y_{\text{R},n}(t) = \sum_{k=1}^K\mathbf{h}_{k,n}\mathbf{x}_{k}(t) + z_{\text{R},n}(t), \ n\in \mathcal{I}_N\triangleq\{1,2,\cdots,N\}, \ t\in \mathcal{I}_{\alpha T}\triangleq\{1,2,\cdots,\alpha T\},
\end{equation}
where $\mathbf{h}_{k,n}\in \mathbb{R}^{1\times M}$ denotes the channel coefficient vector from user $k$ to relay $n$; $\mathbf{x}_{k}(t) \in \mathbb{R}^{M\times 1}$ is the transmitted signal of user $k$ at time $t$; $z_{\text{R},n}(t) \thicksim \mathcal{N}(0, \sigma_{\text{R}}^2)$ is the white Gaussian noise at relay $n$ at time $t$, with $ \sigma_{\text{R}}^2 $ being the noise power at relay. Denote by $\mathbf{Y}_\text{R} \in\mathbb{R}^{N\times \alpha T}$ the received signal at all relays during $\alpha T$ time slots with the $(n, t)$th element of $\mathbf{Y}_\text{R}$ given by $y_{\text{R},n}(t)$. Then, the system model in (\ref{eq.yRn}) can be rewritten into a matrix form:   
\begin{equation}\label{eq.yRmatrix}
	\mathbf{Y}_\text{R} = \sum_{k=1}^K\mathbf{H}_{k} \mathbf{X}_k + \mathbf{Z}_\text{R},
\end{equation}
where $\mathbf{H}_k=[\mathbf{h}_{k,1}^{\text{T}},\mathbf{h}_{k,2}^{\text{T}},\cdots,\mathbf{h}_{k,N}^{\text{T}}]^{\text{T}} \in \mathbb{R}^{N\times M}$ is the channel matrix from user $k$ to all $N$ relays, $ \mathbf{X}_k=[\mathbf{x}_{k}(1),\mathbf{x}_{k}(2),\cdots,\mathbf{x}_{k}(\alpha T)]\in\mathbb{R}^{M\times \alpha T}$ is the transmitted signal from user $ k $ during $ \alpha T $ time slots, and $ \mathbf{Z}_\text{R} \in\mathbb{R}^{N\times \alpha T}$ is the additive noise matrix with the $ (n,t) $th element given by $ z_{\text{R},n}(t) $. The power constraint of user $ k $ in the uplink phase is given by 
\begin{equation}\label{eq.yRmatrixn}
\frac{1}{\alpha T}\mathrm{E}\big[\mathrm{tr}\{\mathbf{X}_k\mathbf{X}_k^{\text{T}} \} \big]\le P_k, \ k\in \mathcal{I}_K\triangleq\{1,2,\cdots,K\},
\end{equation}
where $ P_k $ is the power budget of user $ k $.

In the downlink phase, each relay processes its received signal $ \mathbf{y}_{\text{R}, n} $ as 
\begin{equation}\label{diy1}
	\mathbf{x}_{\text{R},n}=f_{\text{R},n}\big(\mathbf{y}_{\text{R},n}\big),
\end{equation}
where $ \mathbf{y}_{\text{R},n} $ is the $ n $-th row of $ \mathbf{Y}_\text{R} $, $ f_{\text{R},n}(\cdot) $ is the function of relay $ n $, and $ \mathbf{x}_{\text{R},n} \in \mathbb{R}^{1\times (1-\alpha)T} $ represents the signal transmitted by relay $ n $. Then, each relay $ n $ transmits $ \mathbf{x}_{\text{R},n}  $ to the users. The received signal at user $ k $ is given by 
\begin{equation}\label{eq.ykn}
\mathbf{Y}_k = \sum_{n=1}^{N}\mathbf{g}_{n,k}\mathbf{x}_{\text{R},n} + \mathbf{Z}_k, \ k \in \mathcal{I}_K,
\end{equation}
where $ \mathbf{Y}_k\in \mathbb{R}^{M\times (1-\alpha)T} $ denotes the received signal of user $ k $ during $ \alpha T $ time slots, $ \mathbf{g}_{n,k}\in \mathbb{R}^{M\times 1} $ is the channel vector from relay $ n $ to user $ k $,  and $ \mathbf{Z}_k\in \mathbb{R}^{M\times (1-\alpha)T} $ is the white Gaussian noise at user $ k $ with each entry independently drawn from $ \mathcal{N}(0, \sigma_{\text{R}}^2) $. The power constraint of relay $ n $ in the downlink phase is given by 
\begin{equation}\label{eqqrc}
\frac{1}{(1-\alpha)T}\text{E}\big[\text{tr}\{\|\mathbf{x}_{\text{R},n}\|^2_2 \}\big]\le P_{\text{R},n}, \ n\in \mathcal{I}_N,
\end{equation}
where $ P_{\text{R},n} $ is the power budget of relay $ n $. The system model in (\ref{eq.ykn}) can also be written into a matrix form as
\begin{equation}\label{eq.yknmatrix}
\mathbf{Y}_k = \mathbf{G}_{k}\mathbf{X}_{\text{R}} + \mathbf{Z}_k, \ k \in \mathcal{I}_K,
\end{equation}
where $ \mathbf{G}_{k}=[\mathbf{g}_{1,k},\mathbf{g}_{2,k},\cdots,\mathbf{g}_{N,k}]\in \mathbb{R}^{M\times N} $ is the channel matrix from the relays to user $ k $ and $ \mathbf{X}_\text{R}=[\mathbf{x}_{\text{R},1}^{\text{T}},\mathbf{x}_{\text{R},2}^{\text{T}},\cdots,\mathbf{x}_{\text{R},N}^{\text{T}}]^{\text{T}}\in \mathbb{R}^{N\times (1-\alpha)T} $ is the transmitted signal of the relays.

We assume that the elements of $ \mathbf{H}_k $  and $ \mathbf{G}_k $ are independently drawn from a continuous distribution. Then, with probability one, these channel matrices are of full column or row rank, whichever is smaller. We also assume that the channel is block-fading, i.e., the channel remains invariant within each round of data exchange of time duration $ T $. The channel state information is assumed to be \emph{a priori} known to all the nodes in the network, following the convention in \cite{caire99,Jing09}.

\subsection{Achievable Rates}
In the considered MIMO MDRC, each user $ k $, $ k\in \mathcal{I}_K $, broadcasts its messages to all other users. In the uplink phase, user $ k $ broadcasts its message $w_k\in\mathcal{W}_{k}\triangleq\{1,2,\cdots,2^{ TR_{k}}\}$ to other $ K-1 $ users at a rate of $ R_k $; in the downlink phase, user $ k $ estimates messages $ w_{k^\prime}, k^\prime \in \mathcal{I}_K\backslash k$ from other users based on the received signal $ \mathbf{Y}_k $ and the self message $ w_k $. Denote by $\hat{w}_{k,k^\prime}$, the estimate of $w_{k^\prime}$ at user $ k$. Then, the error probability of $w_{k^\prime}$ at user $ k $ is defined as $P_{e, k,k^\prime} \triangleq \Pr \{\hat{w}_{k,k^\prime}\neq w_{k^\prime}\}$. A rate tuple $(R_{1}, \cdots, R_{K})$ is said to be achievable if the error probabilities $\{P_{e, k,k^\prime }| k^\prime \neq k \}$  vanish as $T$ tends to infinity. The capacity region is given by the closure of all possible achievable rate tuples.

\subsection{Capacity Outer Bound}
\begin{figure}[tp]
	\centering
	\includegraphics[width= 0.8\textwidth]{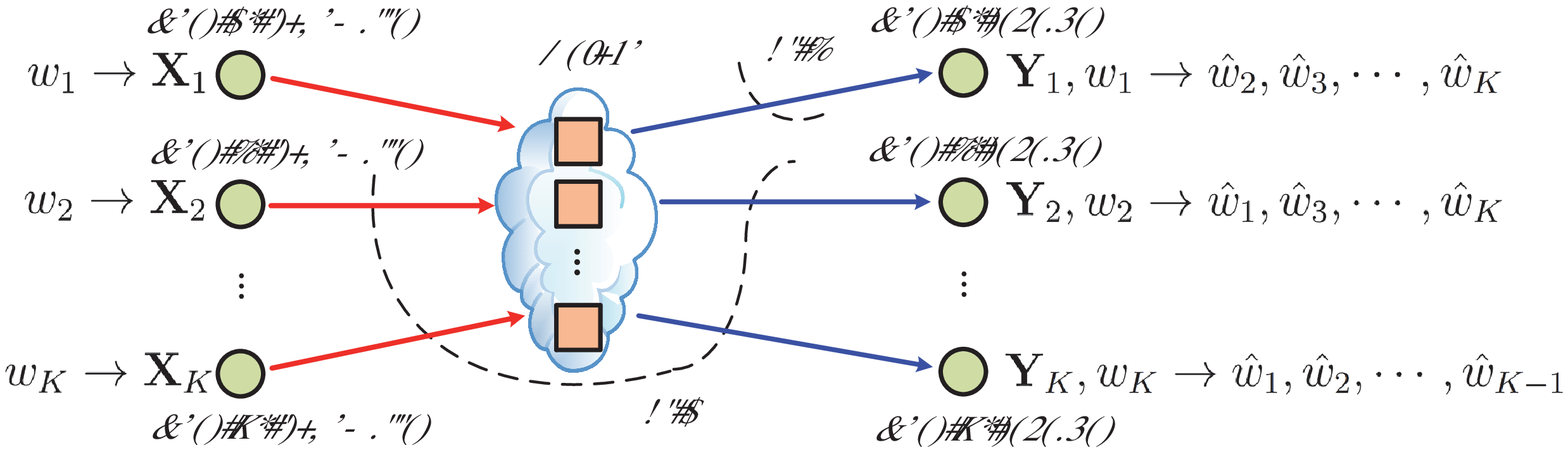}
	\caption{An equivalent system model of the MIMO MDRC with full data exchange and the corresponding cuts for $k = 1$.}
	\label{Figure2}
\end{figure}

We now present a capacity outer bound for the MIMO MDRC. For upper-bound, we assume that relays can fully cooperate with each other, i.e., they form a single super relay. From the cut-set theorem, we obtain 
\begin{subequations}\label{bound}
	\begin{align}
	&\sum_{k^\prime = 1, k^\prime \neq k}^{K}\!\! R_{k^\prime} \leq \frac{\alpha}{2}\log \left\vert \mathbf{I}_{N}+\frac{1}{\sigma_\text{R}^2}\sum_{k^\prime = 1, k^\prime \neq k}^{K}\!\!\mathbf{H}_{k^{\prime}}\mathbf{Q}_{k^\prime}\mathbf{H}_{k^{\prime}}^{\text{T}}\right\vert\label{boundup}\\
	&\sum_{k^\prime = 1, k^\prime \neq k}^{K}\!\! R_{k^\prime} \leq \frac{1-\alpha}{2}\log \left\vert \mathbf{I}_{M}+\frac{1}{\sigma_k^2}\mathbf{G}_{k}\mathbf{Q}_\text{R}\mathbf{G}
	_{k}^{\text{T}}\right\vert, \ \text{for}\ k \in \mathcal{I}_K, \label{bounddown}
	\end{align}
\end{subequations}		
where $\mathbf{Q}_{k}\triangleq \frac{1}{\alpha T}\mathrm{E}\big[\mathbf{X}_{k}\mathbf{X}_{k}^{\text{T}}\big]$ is the covariance matrix of user $ k $, and $\mathbf{Q}_\text{R}\triangleq \frac{1}{(1-\alpha)T}\mathrm{E}\big[\mathbf{X}_\text{R}\mathbf{X}_\text{R}^{\text{T}}\big]$ is the covariance matrix of the super relay. Note that \eqref{boundup} and \eqref{bounddown} are obtained from two different cuts. An illustration of these two cuts for user $ 1 $ (with $ k=1 $) is given in Fig. \ref{Figure2}. For Cut $ 1 $, the information flow is from user $ 2,\cdots,K $ to user $ 1 $. This is a multiple access channel of $ K-1 $ users, with the sum-rate upper-bounded by \eqref{boundup} (with $ k=1 $). Similarly, Cut $ 2 $ also gives a multiple access channel, with the sum-rate upper-bounded by \eqref{bounddown} (with $ k=1 $). Then, a capacity outer bound is given by optimizing $ \{\mathbf{Q}_k, k\in\mathcal{I}_K\}$ and $ \mathbf{Q}_\text{R} $ subject to the following power constraints:
\begin{subequations}
	\begin{eqnarray} 
	&\ \ \ \mathrm{tr}\{\mathbf{Q}_k\}\le P_k,\  k \in \mathcal{I}_K\label{qkc}\\
	&\ \ \mathrm{tr}\{\mathbf{Q}_\text{R}\}\le \sum_{n=1}^{N}P_{\text{R},n},\label{qrc}
	\end{eqnarray}
\end{subequations}
where (\ref{qkc}) is from (\ref{eq.yRmatrixn}), and (\ref{qrc}) is a relaxation of (\ref{eqqrc}). The above optimization problem is convex and can be solved using standard convex programming tools. The goal of this  paper is to develop an efficient communication strategy to approach the outer bound.

\section{Proposed Relay Protocol}\label{sec proposed}

In this section, we consider the transceiver and relay design for the MIMO MDRC with $M \geq N$. We will briefly discuss the case of $M < N$ at the end of the section.

\subsection{Channel Triangularization}
Let the RQ decomposition of $\mathbf{H}_{k}$ be
\begin{equation}\label{hhrr}
	\mathbf{H}_{k} = \mathbf{R}_{k} \mathbf{U}_{k},
\end{equation}
where $\mathbf{R}_{k}\in {\mathbb{R}}^{N\times N}$ is an upper-triangular matrix, and $\mathbf{U}_{k}\in {\mathbb{R}}^{N\times M}$ is a matrix containing the first $ N $ rows of a unitary matrix  satisfying  $\mathbf{U}_{k}\mathbf{U}_{k}^{\text{T}} = \mathbf{I}_{N}$. Then, the received signal at the relays in (\ref{eq.yRmatrix}) can be rewritten as 
\begin{equation}\label{eq.yR2}
	\mathbf{Y}_\text{R}  = \sum_{k=1}^K\mathbf{R}_{k} \widetilde{\mathbf{X}}_k   + \mathbf{Z}_\text{R},
\end{equation}
where $\widetilde{\mathbf{X}}_k  =  \mathbf{U}_{k}\mathbf{X}_k \in \mathbb{R}^{N\times \alpha T} $, and the power constraint in (\ref{eq.yRmatrixn}) is equivalently written as
\begin{equation}\label{i1}
\frac{1}{\alpha T}\mathrm{E}\left[\mathrm{tr}\{\widetilde{\mathbf{X}}_k\widetilde{\mathbf{X}}_k^{\text{T}}\}\right]\le P_k,\ \forall k \in \mathcal{I}_K.
\end{equation}
 We henceforth focus on the transceiver design for the equivalent system given by (\ref{eq.yR2}) and (\ref{eq.ykn}). Each row of $\mathbf{R}_k$ can be seen as a sub-channel. The received signal of the $n$-th sub-channel is given by 
\begin{equation}\label{eq.yR3}
	\mathbf{y}_{\text{R},n} = \sum_{k=1}^K\left(r_{k}(n,n) \widetilde{\mathbf{x}}_{k,n}+\mathbf{v}_{k,n}\right)+ \mathbf{z}_{\text{R},n},
\end{equation}
where $r_{k}(n,n)$ is the $(n,n)$-th element of $\mathbf{R}_{k}$; $\mathbf{y}_{\text{R},n}$, $\widetilde{\mathbf{x}}_{k,n}$, and $\mathbf{z}_{\text{R},n}$ are respectively the $n$-th row of $\mathbf{Y}_\text{R}$, $\widetilde{\mathbf{X}}_k $, and $\mathbf{Z}_\text{R}$; $\mathbf{v}_{k,n}$ is given by
\begin{equation}\label{eq.vkn}
	\mathbf{v}_{k,n} = \sum_{n^\prime = n+1}^{N}r_{k}(n,n^\prime) \widetilde{\mathbf{x}}_{k,n^\prime}.
\end{equation}
Note that $\mathbf{y}_{\text{R},n}$ in \eqref{eq.yR3} is the received signal at relay $n$. In the following, we describe encoding and decoding operations based on the system model given by \eqref{eq.yR3} and \eqref{eq.ykn}.

\subsection{Uplink Phase: User Encoding}
\begin{figure}[!t]
	\centering
	\includegraphics[width=0.6\textwidth]{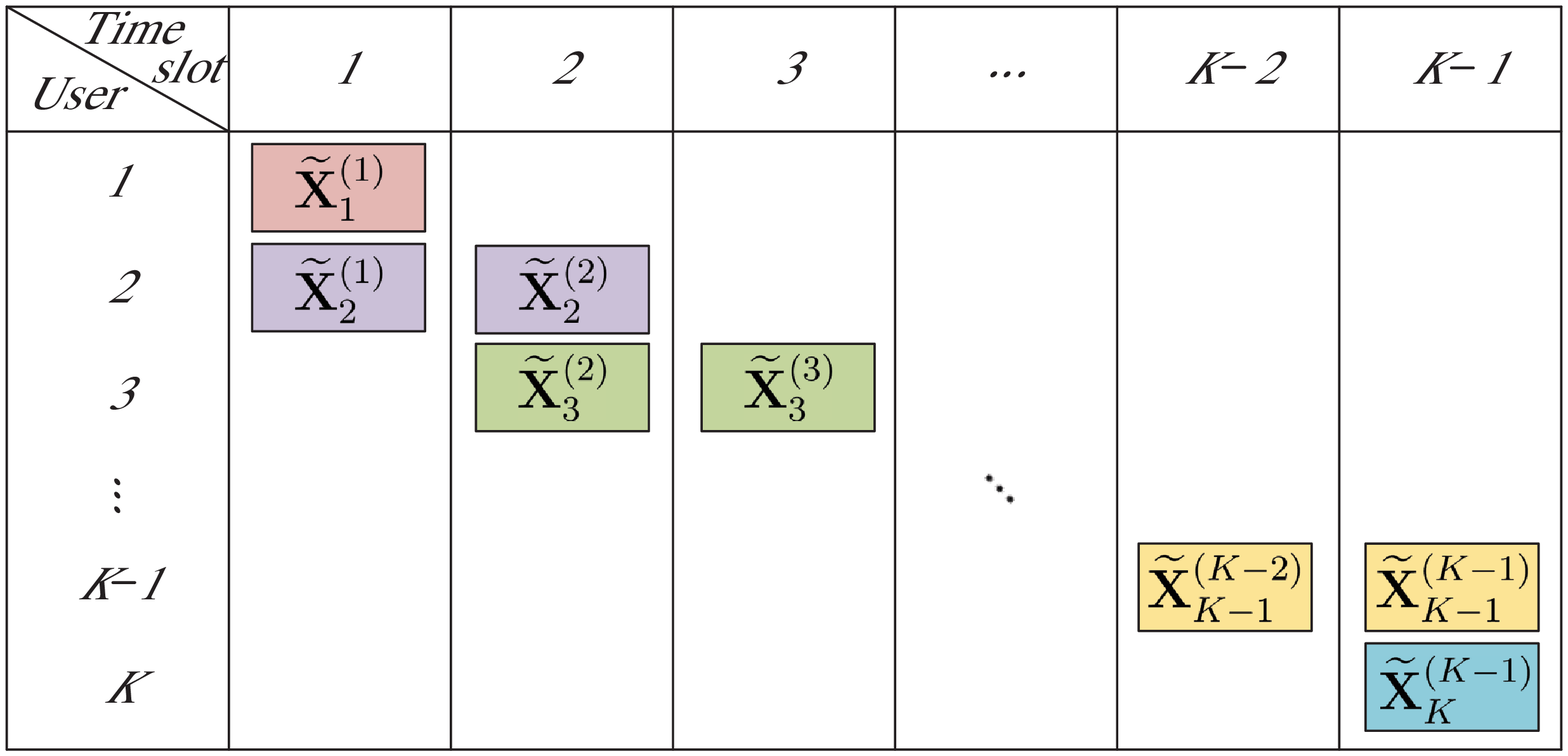}
	\caption{Uplink user scheduling, where $\widetilde{\mathbf{x}}_k^{(l)} = \left[\widetilde{\mathbf{x}}_k^{(l)}(1)^{\text{T}}, \cdots, \widetilde{\mathbf{x}}_k^{(l)}(N)^{\text{T}}\right]^{\text{T}}$, and $\widetilde{\mathbf{x}}_{k+1}^{(k)} = \widetilde{\mathbf{x}}_{k+1}^{(k+1)}$ for $k = 1, \cdots, K-2$.}
	\label{Figure3}
\end{figure}
 	In this subsection, we present the encoding operations at user ends in the uplink phase. We divide the uplink phase into $K-1$ time slots with equal duration $T^\prime = \frac{\alpha T}{K-1}$.  Let $\mathbf{y}_{\text{R},n}^{(l)}$, $\widetilde{\mathbf{x}}_{k,n}^{(l)}$, $\mathbf{v}_{k,n}^{(l)}$, and $\mathbf{z}_{\text{R},n}^{(l)}$ be the corresponding length-$T^\prime$ signal components of $\mathbf{y}_{\text{R},n}$, $\widetilde{\mathbf{x}}_{k,n}$, $\mathbf{v}_{k,n}$, and $\mathbf{z}_{\text{R},n}$, in the $l$-th time slot. That is,
\begin{subequations}
	\begin{align}\label{eq.def}
		\mathbf{y}_{\text{R},n} &= [\mathbf{y}_{\text{R},n}^{(1)}, \cdots, \mathbf{y}_{\text{R},n}^{(K-1)}]\\
		\widetilde{\mathbf{x}}_{k,n} &= [\widetilde{\mathbf{x}}_{k,n}^{(1)}, \cdots, \widetilde{\mathbf{x}}_{k,n}^{(K-1)}]\\
		\mathbf{v}_{k,n} &= [\mathbf{v}_{k,n}^{(1)}, \cdots, \mathbf{v}_{k,n}^{(K-1)}]\\
		\mathbf{z}_{\text{R},n} &= [\mathbf{z}_{\text{R},n}^{(1)}, \cdots, \mathbf{z}_{\text{R},n}^{(K-1)}].
	\end{align}
\end{subequations}
From \eqref{eq.yR3}, the received signal vector of the $n$-th sub-channel in the $l$-th time slot is given by
\begin{equation}\label{eq.yR33}
	\mathbf{y}_{\text{R},n}^{(l)} = \sum_{k=1}^K\left(r_{k}(n,n) \widetilde{\mathbf{x}}_{k,n}^{(l)}+\mathbf{v}_{k,n}^{(l)}\right) + \mathbf{z}_{\text{R},n}^{(l)}.
\end{equation}

Let ${\cal{W}}_{k,n} = \{1,2,\cdots,2^{TR_{k,n}} \}$ be the message set for the spatial data stream of the $k$-th user over the $n$-th sub-channel, and $w_{k,n} \in {\cal{W}}_{k,n}$ be the corresponding message, where $R_{k,n}$ is the information rate of user $k$ over the $n$-th sub-channel. We refer to $\{w_{1,n}, \cdots, w_{K,n}\}$ as the message tuple of the sub-channel $n$.

Nested lattice coding \cite{Erez04}, \cite{Ordentlich13} is applied to each message tuple $\{w_{1,n}, \cdots, w_{K,n}\}$. The codebooks for the $n$-th sub-channel are constructed as follows. Without loss of generality, let $\pi(\cdot)$ be the permutation with the permuted indices $\{\pi(1),\cdots,\pi(K)\}$ satisfying $R_{\pi(1),n}\geq R_{\pi(2),n}\geq\cdots\geq R_{\pi(K),n}$. We construct a chain of \textquotedblleft good\textquotedblright\ nested lattices $\Lambda_{\pi(1),n}$, $\cdots$, $\Lambda_{\pi(K),n}$, and $\Lambda_{C,n}$ satisfying $\Lambda_{\pi(1),n}  \subseteq \cdots \subseteq \Lambda_{\pi(K),n} \subseteq \Lambda_{C,n}\in \mathbb{R}^{T^\prime}$ \cite{Ordentlich13}. Let $\mathcal{C}_{k,n}$ be the nested lattice code defined by $\Lambda_{k,n}$ and $\Lambda_{C,n}$, and $\mathbf{c}_{k,n} \in \mathcal{C}_{k,n}$ be the codeword mapped from the message $w_{k,n}$.\footnote{Note that the codeword length of each user for each sub-channel is $ T^\prime $ (instead of $ \alpha T $). This implies that a codeword may be repeatedly transmitted over a sub-channel by at most $ \alpha T/T^\prime=K-1 $ times.}
For indexes $k$ and $k^\prime$, $\Lambda_{k,n}$ and $\Lambda_{k^\prime,n}$ are constructed with the nesting ratio satisfying
\begin{equation}
	\log\left(\frac{\text{Vol}(\Lambda_{k,n})}{\text{Vol}(\Lambda_{k^\prime,n})}\right)  =  \frac{T^\prime}{2}\log\left(\frac{|r_{k}(n,n)|^2 P_{k,n}} {|r_{{k^\prime}R}(n,n)|^2 P_{k^\prime,n}}\right) + O(1),\nonumber
\end{equation}
where $\text{Vol}(\Lambda)$ is the the volume of the fundamental Voronoi region of a
lattice $\Lambda$, $P_{k,n}$ denotes the average power of the $n$-th spatial stream of user $k$, and  $O(1) $ is bounded  as $T \to \infty$. Then, the relation between $R_{k,n}$ and $R_{k^\prime,n}$ can be written as
\begin{equation}\label{eq.ratebm}
	R_{k,n} = R_{k^\prime,n} + \frac{\alpha}{2(K-1)}\log \left( \frac{|r_{k}(n,n)|^2 P_{k,n}} {|r_{{k^\prime}R}(n,n)|^2 P_{k^\prime,n}}\right) + o\Big(\frac{1}{T}\Big).
\end{equation}

The encoding of each user $k$ is ordered reversely from the $N$-th sub-channel to the first sub-channel. 
For each sub-channel $n$, let $\mathbf{d}_{k,n}$ be a random dithering vector that is uniformly distributed over the Voronoi region of $\Lambda_{k,n}$. The dithering signals are globally known to all the nodes in the network. With dithering, the $n$-th transmit signal $\widetilde{\mathbf{x}}^{(l)}_{k,n}$ of user $k$ at time slot $l$ is constructed by
\begin{eqnarray}\label{i2}
\widetilde{\mathbf{x}}^{(l)}_{k,n} =
\begin{cases}
 \frac{1}{r_{k}(n,n)}\left(\left(\mathbf{c}_{k,n}\! -\! \mathbf{v}^{(l)}_{k,n}\! -\! \mathbf{d}_{k,n} \right)~\text{mod}~\Lambda_{k,n}\right),\ \ \text{for}\ k=l,l+1;\\
0,\ \ \text{for}\ k\neq l,k\neq l+1.
\end{cases}
\end{eqnarray}
In \eqref{i2}, only user $ l $ and $ l+1 $ are active in each time slot $ l $.\footnote{It can be shown that with nested lattice coding, concurrent signaling of more than two users in a time slot generally leads to power inefficiency.} The uplink user scheduling is illustrated in Fig. \ref{Figure3}.  Given the encoding order at each user $ k $, the inter-stream interference $\mathbf{v}^{(l)}_{k,n}$ defined in \eqref{eq.vkn} is \emph{a priori} known by user $k$ when the $n$-th spatial stream of user $k$ is encoded. Therefore, $ \widetilde{\mathbf{x}}_{k,n}^{(l)} $ in \eqref{i2} is indeed constructed. With \eqref{i2}, the received signal in \eqref{eq.yR33} reduces to 
\begin{equation}\label{eq.yR34}
	\mathbf{y}_{\text{R},n}^{(l)} = \sum_{k=l}^{l+1}\left(r_{k}(n,n) \widetilde{\mathbf{x}}_{k,n}^{(l)}+\mathbf{v}_{k,n}^{(l)}\right) + \mathbf{z}_{\text{R},n}^{(l)}.
\end{equation}

From Fig. \ref{Figure3}, we see that each user $k$ transmits signals only at time slots $k-1$ and $k$, except that user $1$ is active only at time slot $1$ and user $K$ is active only at time slot $K-1$. Also, for each user $k \in \{2, \cdots, K-1\}$, the signals transmitted over the two time slots are identical to each other.  Thus, we have
\begin{equation}
	\widetilde{\mathbf{x}}_{k+1,n}^{(k)}= \widetilde{\mathbf{x}}_{k+1,n}^{(k+1)},\ \mathbf{v}_{k+1,n}^{(k)} = \mathbf{v}_{k+1,n}^{(k+1)}, \text{ for }k = 1, \cdots, K-2.
\end{equation}


\subsection{Relay Operations}
We now consider relay decoding. At time slot $ l $, the received signal at relay $ n $ is given by 
\begin{equation} \label{i3}
	\mathbf{y}_{\text{R},n}^{(l)} = \sum_{k=l}^{l+1}\left(r_{k}(n,n)\widetilde{\mathbf{x}}^{(l)}_{k,n} +\mathbf{v}^{(l)}_{k,n}\right)+\mathbf{z}^{(l)}_{k,n},\ l\in \mathcal{I}_{K-1},\ n\in \mathcal{I}_N.
\end{equation}
Without loss of generality, we assume $ \Lambda_{l,n}\subseteq \Lambda_{l+1,n}$. Then, upon receiving $ \mathbf{y}_{\text{R},n}^{(l)}$, relay $ n $ computes 
\begin{subequations}\label{eq.der1}
\begin{align}
\widetilde{\mathbf{y}}_{\text{R},n}^{(l)} &= \bigg(\mathbf{y}_{\text{R},n}^{(l)} + \sum_{k=l}^{l+1}\mathbf{d}_{k,n}\bigg) \text{mod}\  \Lambda_{l,n}\label{dda}\\
&=\bigg(\sum_{k=l}^{l+1}\Big[\big(\mathbf{c}_{k,n}-\mathbf{v}^{(l)}_{k,n}-\mathbf{d}_{k,n}\big)\text{mod}\ \Lambda_{k,n}+\mathbf{v}^{(l)}_{k,n}\Big]+ \mathbf{z}_{\text{R},n}^{(l)}+\sum_{k=l}^{l+1}\mathbf{d}_{k,n}\bigg) \text{mod}\  \Lambda_{l,n}\label{ddb}\\
&=\bigg(\sum_{k=l}^{l+1}\Big(\mathbf{c}_{k,n}-\mathbf{Q}_{\Lambda_{k,n}}\big(\mathbf{c}_{k,n}+\mathbf{v}^{(l)}_{k,n}+\mathbf{d}_{k,n}\big)\Big)+\mathbf{z}_{\text{R},n}^{(l)} \bigg) \text{mod}\  \Lambda_{l,n}  \label{ddc}\\
&= \Big(\mathbf{w}_{\text{R},n}^{(l)}+\widetilde{\mathbf{z}}_{\text{R},n}^{(l)}\Big)\text{ mod}\  \Lambda_{l,n},\label{ddd}
\end{align}
\end{subequations}
where
\begin{subequations}
\begin{align}
\mathbf{w}_{\text{R},n}^{(l)} &= \big(\mathbf{c}_{l,n}+\mathbf{c}_{l+1,n}\big)\text{ mod } \Lambda_{l,n}\label{xxa}\\
\widetilde{\mathbf{z}}_{\text{R},n}^{(l)}&= \Big(\mathbf{z}_{\text{R},n}^{(l)} - \mathbf{Q}_{\Lambda_{l+1,n}}\big(\mathbf{c}_{l+1,n}+\mathbf{v}^{(l)}_{l+1}(n)+\mathbf{d}_{l+1,n}\big)\Big) \text{ mod }  \Lambda_{l,n},
 \end{align}
\end{subequations}
 and $ \mathbf{Q}_{\Lambda}(\cdot) $ denotes the lattice quantizer that outputs the lattice point of $ \Lambda $ closest to the input. Note that \eqref{ddb} follows from \eqref{i2} and \eqref{i3}, \eqref{ddc} follows from $ \mathbf{x} \text{ mod}\  \Lambda =\mathbf{x}-\mathbf{Q}_{\Lambda}(\mathbf{x}) $, and \eqref{ddd} utilizes $ (\mathbf{x}+\mathbf{y})\text{ mod}\ \Lambda =\big((\mathbf{x}\ \text{mod}\  \Lambda )+(\mathbf{y}\ \text{mod}\  \Lambda )\big)\ \text{mod}\  \Lambda $. Both $ \mathbf{c}_{l,n} $ and $ \mathbf{c}_{l+1,n} $ are lattice points of $ \Lambda_{C,n} $, and so is $\mathbf{w}_{\text{R},n}^{(l)}$. Each relay wants to decode the combination $ \mathbf{w}_{\text{R},n}^{(l)} $. Let $ \hat{\mathbf{w}}_{\text{R},n}^{(l)} $ be an estimate of $ \mathbf{w}_{\text{R},n}^{(l)} $ given $ \mathbf{y}_{\text{R},n}^{(l)}$. With ambiguity decoding \cite{Erez04}, the probability of $\hat{\mathbf{w}}_{\text{R},n}^{(l)} \neq \mathbf{w}_{\text{R},n}^{(l)}$ vanishes as $T \to \infty$ provided that
\begin{equation}\label{eq.ulratebr}
R_{k,n} \le \frac{1}{K-1}\left[ \frac{\alpha }{2}\log\left( \frac{|r_{k}(n,n)|^2 P_{k,n} }{\sigma_{\text{R}}^2}\right) \right]^+,\ \ \text{for}\ k \in \mathcal{I}_K, n\in\mathcal{I}_N.
\end{equation}
Then, each relay $ n $ re-encodes the estimated messages $\{\hat{\mathbf{w}}_{\text{R},n}^{(l)} |  l = 1,\cdots, K-1\}$ to obtain the transmitted signal $ \mathbf{x}_{\text{R},n} $ and broadcasts $ \mathbf{x}_{\text{R},n} $ to the users. The encoding function of each relay is described as follows. Let $ \widetilde{R}_n $ be the total rate of $\{\hat{\mathbf{w}}_{\text{R},n}^{(l)} |  l = 1,\cdots, K-1\}$. Each relay $ n ,n\in \mathcal{I}_N$, constructs a random codebook ${\cal{C}}_{\text{R},n}$ with cardinality $2^{T\widetilde{R}_n}$. The encoding function of relay $ n $, denoted by $ f_{\text{R},n} $, is to map every  message tuple $\big(\hat{\mathbf{w}}_{\text{R},n}^{(1)},\cdots,\hat{\mathbf{w}}_{\text{R},n}^{(K-1)}\big)$ to a unique codeword in $ \mathcal{C}_{\text{R},n} $. Then, the transmitted signal at relay $ n $ is given by $\mathbf{x}_{\text{R},n} = f_{\text{R},n}(\{\hat{\mathbf{w}}_{\text{R},n}^{(l)} | l=1,\cdots,K-1\})$. 

\subsection{Downlink Phase: User Decoding}
We now focus on the decoding operation at the user side. From (\ref{eq.ykn}), the received signal at each user $ k $ is given by 
\begin{equation}\label{eq.ykn2}
\mathbf{Y}_k = \sum_{n=1}^{N}\mathbf{g}_{n,k}\mathbf{x}_{\text{R},n} + \mathbf{Z}_k, \ k \in \mathcal{I}_K, n\in\mathcal{I}_N.
\end{equation}

Upon receiving $ \mathbf{Y}_k $, each user $ k $ wants to learn $ \{\mathbf{c}_{k,n}| n\in \mathcal{I}_N, k^\prime\in \mathcal{I}_K\backslash k\}$ with the help of the self message $ \{\mathbf{c}_{k,n}| n\in \mathcal{I}_N\} $. To this end, each user $ k $ takes a two-step procedure as follows:\\
\emph{Step 1):} Decode $\{\hat{w}_{k^\prime,n}|n\in\mathcal{I}_N,k^\prime\in\mathcal{I}_K\backslash k\} $ based on $ \mathbf{Y}_k $ and $ \{\mathbf{c}_{k,n}| n\in \mathcal{I}_N\}$.\\
\emph{Step 2):} Retrieve $ \{\mathbf{c}_{k,n}| n\in \mathcal{I}_N, k^\prime\in \mathcal{I}_K\backslash k\}$ from $ \{\hat{w}_{k^\prime,n}| n\in\mathcal{I}_N,k^\prime =1,\cdots,K-1 \}$ and $\{\mathbf{c}_{k,n}| n\in \mathcal{I}_N\} $.

We first consider Step 1. Note that, given $\{\mathbf{c}_{k,n}| n\in \mathcal{I}_N\} $, the rate of $ \{\hat{w}_{k^\prime,n}| k^\prime =1,\cdots,K-1 \}$ (i.e., the message tuple of relay $ n $) is reduced to $ \sum_{k^\prime\neq k}^{K} R_{k^\prime,n}$. Therefore, the channel model in \eqref{eq.ykn2} becomes an $ N $-terminal multiple access channel with the transmission rates of the $ N $ terminals given by $ \{\sum_{k^\prime\neq k}^{K} R_{k^\prime,n}|n\in\mathcal{I}_N\}$. The corresponding capacity region is given by
\begin{equation}\label{eq.dlratem}
\sum_{n\in \mathcal{S}}\sum_{k^\prime\in \mathcal{I}_K\backslash k}R_{k^\prime,n} \leq \!\frac{1-\alpha }{2}\log\bigg| \mathbf{I}_M +\frac{1}{\sigma^2_k} \sum_{n\in \mathcal{S}} P_{\text{R},n} \mathbf{g}_{n,k} \mathbf{g}_{n,k}^{T} \bigg |,\ \ \forall k \in \mathcal{I}_K,\forall \mathcal{S} \subseteq \mathcal{I}_N,
\end{equation}
where $ \mathcal{S} $ is an arbitrary subset of $ \mathcal{I}_N $.

We now consider Step 2. First note that $ \hat{\mathbf{w}}_{R,n}^{(l)}=\mathbf{w}_{R,n}^{(l)} $ for any $ l $ and $ n $ provided that \eqref{eq.ulratebr} holds. Then, it suffices to show that, for any given $ k $ and $ n $, $\{\mathbf{c}_{k^\prime,n}|k^\prime\in\mathcal{I}_K\backslash k\} $ can be retrieved from $\{\hat{\mathbf{w}}_{\text{R},n}^{(l)} |  l = 1,\cdots, K-1\}$ and $ \mathbf{c}_{k,n} $. This is true from the definition of $ \mathbf{w}_{R,n}^{(l)} $ in \eqref{xxa}.

\subsection{Achievable Rate of the Overall Scheme}
We summarize the  discussions in the preceding subsections as a theorem below.
\begin{theorem} A rate tuple $(R_{1},\ldots, R_{K})$  is achievable for the considered MIMO MDRC if
	\begin{subequations}\label{23}
		\begin{align}
			R_{k,n}\! &\leq \!\frac{1}{K\!-\!1}\!\left[ \frac{\alpha }{2}\log\bigg( \frac{|r_{k}(n,n)|^2 P_{k,n} }{\sigma_{\text{R}}^2}\bigg) \right]^+\!\!, k \in I_K\quad \ \label{20a}\\
			\sum_{n\in \mathcal{S}}\sum_{k^\prime\in \mathcal{I}_K\backslash k}R_{k^\prime,n} &\leq \!\frac{1-\alpha }{2}\log\left| \mathbf{I}_M +\frac{1}{\sigma^2_k}\sum_{n\in \mathcal{S}}P_{\text{R},n}\mathbf{g}_{n,k}\mathbf{g}_{n,k}^{T} \right |,\ \ k \in \mathcal{I}_K, \mathcal{S}\subseteq \mathcal{I}_N,\ \ \label{20b}
		\end{align}
		where $P_{k,n}$ is the transmission power of the $n$-th sub-channel of user $k$ satisfying
		\begin{align}
			&\frac{2}{K-1}\sum_{n=1}^{N}P_{k,n} \leq P_k,\  \mathrm{for}\ k = 2, \cdots, K-1\ \ \label{20c}\\
			&\frac{1}{K-1}\sum_{n=1}^{N}P_{k,n} \leq P_k,\  \mathrm{for}\ k = 1,K \label{20d}.
		\end{align}
	\end{subequations}
\end{theorem}
\subsection{Further Discussions}
For comparison, we consider cooperative relaying in which the relays are combined as a super node with $ N $ antennas. In this case, upon receiving the signal $\{\mathbf{y}_{\text{R},n}^{(l)} | l \in \mathcal{I}_{K-1},\ n \in \mathcal{I}_N\}$, the super node (relay) obtains an estimate of $\{\mathbf{w}_{\text{R},n}^{(l)} | l \in \mathcal{I}_{K-1},\ n\in \mathcal{I}_N\}$, denoted by  $\{\hat{\mathbf{w}}_{\text{R},n}^{(l)} | l \in \mathcal{I}_{K-1},\ n\in \mathcal{I}_N\}$. Then, the super node constructs a codebook $\mathcal{C}_\text{R}$ of cardinality $ 2^{T R_{\text{sum}}} $ with $ R_{\text{sum}}\triangleq \sum_{k=1}^{K}\sum_{n=1}^{N}R_{k,n} $. Each codeword is an $ N $-by-$ (1-\alpha)T $ random matrix with each column independently drawn from $ \mathcal{N}(\mathbf{0},\mathbf{Q}_\text{R}) $, where $ \mathbf{0} $ is an all-zero vector with an appropriate size, and $ \mathbf{Q}_\text{R} $ is the covariance matrix at the super-relay satisfying the total power constraint in \eqref{qrc}. The codeword of the message tuple $\{\hat{\mathbf{w}}_{\text{R},n}^{(l)}\}$ is then broadcast to all users in the downlink phase. The achievable rate tuple satisfies the following inequalities:
\begin{subequations}\label{crate}
	\begin{align}
	&R_{k} \leq \frac{1}{K-1}\sum_{n=1}^{N}\left[ \frac{\alpha }{2}\log\left( \frac{|r_{k}(n,n)|^2 P_{k,n} }{\sigma_\text{R}^2}\right) \right]^+,\ k \in \mathcal{I}_K\label{crateup} \\
	&\sum_{k^\prime\in \mathcal{I}_K\backslash k}^{K}R_{k^\prime} \leq \frac{1-\alpha }{2}\log \left\vert \mathbf{I}_{M}+\frac{1}{\sigma_k^2}\mathbf{G}_{k}\mathbf{Q}_{\text{R}}\mathbf{G}\label{cratedown}
	_{k}^{\text{T}}\right\vert, \ k \in \mathcal{I}_K,
	\end{align}
\end{subequations}	
where $ \mathbf{Q}_\text{R} $ satisfies the power constraint in \eqref{qrc}, and $\{P_{k,n}\}$ are constrained by  \eqref{20c} and \eqref{20d}. The result in \eqref{crate} serves as a performance upper bound for the case of distributive relays.

Before leaving this section, we note that $ M\ge N $ is assumed throughout the paper. The proposed scheme can be straightforwardly extended to the case of $M < N$ by disabling $N-M$ relay antennas. This antenna disablement approach is simple but generally not efficient. To improve efficiency, we can reduce the signal space seen at the relay by projecting the received signal $\mathbf{Y}_\text{R}$ into an appropriately chosen subspace of dimension $M$, similarly to the reduced-dimension approach in \cite{Yuanxj2012}. Then, the proposed scheme is applicable to the reduced network. Nevertheless, the detailed design of the projection is out of the scope of this paper.

\section{Sum-Rate Maximization}\label{sec sumrate}
Based on Theorem 1, the sum-rate maximization problem of the proposed lattice coding scheme with distributed relays is formulated as
\begin{subequations} \label{equ:LP-Dist}
	\begin{align}		
		\mathop{\text{maximize}}\limits_{\{P_{k,n}\}, \{R_{k,n}\} }\ \ \ \ &\sum_{k=\mathrm{1}}^{K} \sum_{n = 1}^N R_{k,n}\\
		\text{subject to}\ \ \ \
		&R_{k,n}\! \le \!\frac{1}{K\!-\!1}\!\left[ \frac{\alpha }{2}\log\left( \frac{|r_{k}(n,n)|^2 P_{k,n} }{\sigma_{\text{R}}^2}\right) \right]^+\!\!, k \in \mathcal{I}_K, n\in \mathcal{I}_N \label{equ:LP-Dist-b} \\
		&\sum_{n\in \mathcal{S}}\sum_{k^\prime\in \mathcal{I}_K\backslash k}R_{k^\prime,n} \leq \!\frac{1-\alpha }{2}\log\left\vert\mathbf{I}_M \!+\!\frac{1}{\sigma^2_k}\sum_{n\in \mathcal{S}}P_{\text{R},n}\mathbf{g}_{n,k}\mathbf{g}_{n,k}^{T} \right \vert,\ \ k \in \mathcal{I}_K, \mathcal{S}\subseteq \mathcal{I}_N\label{equ:LP-Dist-c} \\
		&\frac{2}{K-1}\sum_{n=1}^{N}P_{k,n} \leq P_k,\ \ \mathrm{for}\ k = 2, \cdots, K-1 \\
		&\frac{1}{K-1}\sum_{n=1}^{N}P_{k,n} \leq P_k,\ \ \mathrm{for}\ k = 1\ \mathrm{or}\ K.
	\end{align}
\end{subequations}
The above problem is not a convex problem due to the $ [\cdot]^+ $ operator in \eqref{equ:LP-Dist-b}. It can be solved by following the idea of iterative water-filling: First solve \eqref{equ:LP-Dist} using convex programming by removing all the $ [\cdot]^+ $ operators in \eqref{equ:LP-Dist-b}; then construct an index set $ \mathcal{T} : (k,n)\in \mathcal{T}$ if $ R_{k,n} \le 0$ in the solution of the previous step; fix $ P_{k,n}=0$ for $(k,n)\in \mathcal{T} $ and solve \eqref{equ:LP-Dist} again using convex programming. Repeat the above process until there is no new $ (k,n) $ satisfying $ R_{k,n} <0 $.  The above algorithm is guaranteed to converge as the number of deactivated channels (i.e. the cardinality of $ \mathcal{T}$) monotonically increases in iteration.

For comparison, we now describe the sum-rate maximization problem for the case of cooperative relays. From \eqref{crate} and the discussions therein, the corresponding sum-rate maximization problem is given by
\begin{subequations} \label{equ:LP-Coop}
	\begin{align}		
	\mathop {\text{maximize}}\limits_{\mathbf{Q}_\text{R},\{P_{k,n}\},\{R_k\} }\ \ \ \ &\sum_{k=\mathrm{1}}^{K}R_k \\
	\text{subject to}\ \ \ \
	&R_{k} \le \frac{1}{K-1}\sum_{n=1}^{N}\left[ \frac{\alpha }{2}\log\left( \frac{|r_{k}(n,n)|^2 P_{k,n} }{\sigma_\text{R}^2}\right) \right]^+, k\in \mathcal{I}_K \label{equ:LP-Coop-b} \\
	&\sum_{k^\prime\in \mathcal{I}_K\backslash k}^{K}R_{k^\prime} \leq \frac{1-\alpha }{2}\log \left\vert \mathbf{I}_{M}+\frac{1}{\sigma_k^2}\mathbf{G}_{k}\mathbf{Q}_\text{R}\mathbf{G}
	_{k}^{\text{T}}\right\vert, \ k\in\mathcal{I}_K \label{equ:LP-Coop-c} \\
	&\mathrm{tr} \{\mathbf{Q}_\text{R}\}\leq \sum_{n=1}^{N}P_{\text{R},n},  \mathbf{Q}_\text{R}\succeq\mathbf{0} \label{equ:LP-Coop-d} \\
	&\frac{2}{K-1}\sum_{n=1}^{N}P_{k,n} \leq P_k,\ \ \mathrm{for}\ k = 2, \cdots, K-1 \label{equ:LP-Coop-e}\\
	&\frac{1}{K-1}\sum_{n=1}^{N}P_{k,n} \leq P_k,\ \ \mathrm{for}\ k = 1\ \mathrm{or}\ K.\label{equ:LP-Coop-f}
	\end{align}
\end{subequations}			
Compared with \eqref{equ:LP-Dist}, the main difference of \eqref{equ:LP-Coop} is that the downlink rate constraint \eqref{equ:LP-Dist-c} is replaced by \eqref{equ:LP-Coop-c}. Similarly to \eqref{equ:LP-Dist}, the problem in \eqref{equ:LP-Coop} can be solved in an iterative fashion. We omit the details for brevity.
\newcounter{mytempeqncnt}

\section{Asymptotic Analysis}\label{sec asy}
In this section, we analyze the asymptotic behavior of the proposed scheme in the high SNR region. For convenience of discussion, we assume the following settings:  $P_k = \beta_k P,\ k\in \mathcal{I}_K$,  $P_{\text{R},n} = \beta_{\text{R}} P, \ n\in \mathcal{I}_N$, and $\sigma^2_k = \sigma_\text{R}^2 =\sigma^2, \forall k\in \mathcal{I}_K$, where $\{\beta_k\}$ and $\beta_{\text{R}}$ are constants as $P/\sigma^2 \rightarrow \infty $. We first show that, as $P/\sigma^2 \rightarrow \infty $, the sum-rate gap between the proposed scheme with distributive relays and the scheme with cooperative relays tends to zero. Then, we discuss the asymptotic optimality of the proposed scheme by comparison with the cut-set upper bound.

\subsection{Distributive Relaying vs. Cooperative Relaying}
Let $R_{\mathrm{dist}}$ be the optimal sum-rate of the proposed distributive relaying scheme given by \eqref{equ:LP-Dist}, and $R_{\mathrm{coop}}$ be the corresponding sum-rate given by \eqref{equ:LP-Coop}.
\begin{theorem} \label{theorem_DC}
	 When $ \alpha\in (0,\frac{1}{2})\cup(\frac{1}{2},1)$, the proposed distributive relaying scheme achieves the same sum-rate as the cooperative relaying scheme in the high SNR regime, i.e.
	 $$\lim_{P/\sigma^2 \rightarrow \infty} \big(R_{\mathrm{coop}}-R_{\mathrm{dist}}\big)=0.$$
\end{theorem}

\textit{Proof:}
We first consider the distributive relaying scheme with the rate given by \eqref{equ:LP-Dist}. In general, a rate $ R $ is a function of the transmission power $ P $, denoted by $ R(P) $. Then, the corresponding DoF is defined by 
	\begin{equation}
	d=\lim_{P\rightarrow \infty} \frac{R(P)}{P}.
	\end{equation}
 Denote by $ d^{\text{dist}}_{k,n} $ the DoF of use $ k $ over sub-channel $ n $, where the ``dist" is an abbreviation of distributive. Then, from \eqref{equ:LP-Dist-b}, in the uplink phase, we have 
\begin{subequations}\label{dof_up_dist}
	\begin{align}	
	d^{\text{dist}}_{k,n}&\le \lim_{P\rightarrow \infty} \frac{\frac{1}{K\!-\!1}\!\left[ \frac{\alpha }{2}\log\left( \frac{|r_{k}(n,n)|^2  \beta_{k,n} P }{\sigma_{\text{R}}^2}\right) \right]^+}{\log\big( \beta_{k,n} P\big)}\\
	&=\frac{\alpha}{2}\cdot\frac{1}{K-1}, \ \ k\in\mathcal{I}_K, n\in \mathcal{I}_N,
	\end{align}
\end{subequations}
where $ P_{k,,n}=\beta_{k,n}P $. Denote by $ \mathcal{D}^{\text{dist}}_{\text{up}} $ the uplink DoF region specified by \eqref{dof_up_dist}. Clearly, $ \mathcal{D}^{\text{dist}}_{\text{up}} $ is a polyhedron. From \eqref{equ:LP-Dist-c}, in the downlink phase, we have 	
\begin{subequations}\label{dof_down_dist}
\begin{align}	
\sum_{n\in \mathcal{S}}\sum_{k^\prime\in \mathcal{I}_K\backslash k}d^{\text{dist}}_{k^\prime,n} &\le \lim_{P\rightarrow \infty}\frac{\frac{1-\alpha }{2}\log\left\vert\mathbf{I}_M \!\!+\!\!\frac{1}{\sigma^2_k}\sum_{n\in \mathcal{S}}^{}\beta_\text{R} P\mathbf{g}_{n,k}\mathbf{g}_{n,k}^{T} \right \vert}{\log\big( \beta_\text{R} P\big)} \\
&=|\mathcal{S}|\cdot\frac{1-\alpha}{2}, \ \ k\in\mathcal{I}_K, \mathcal{S}\subseteq \mathcal{I}_N,\label{dof_down_dist-b}
\end{align}
\end{subequations}
where $ |\mathcal{S}| $ represents the cardinality of set $ \mathcal{S} $, and \eqref{dof_down_dist-b} utilizes the fact that  $ |\mathcal{S}|\le N\le M $. In  \eqref{dof_down_dist}, the inequalities with $ |\mathcal{S}|=1 $ imply all the other inequalities. Thus, \eqref{dof_down_dist} can be simplified as 
\begin{equation}\label{dof_down_dist1}
\begin{split}	
\sum_{k^\prime\in \mathcal{I}_K\backslash k}d^{\text{dist}}_{k^\prime,n} \le\frac{1-\alpha}{2}, \ \ k\in\mathcal{I}_K, n\in \mathcal{I}_N.
\end{split}
\end{equation}
Denote by  $ \mathcal{D}^{\text{dist}}_{\text{down}} $ the downlink DoF region given by  \eqref{dof_down_dist1}.

We now consider the cooperative relaying scheme with the rates given by \eqref{equ:LP-Coop}. Denote by $ d^{\text{coop}}_{k} $  the corresponding  DoF of user $ k $. Then, from \eqref{equ:LP-Coop-b}, 
\begin{subequations}\label{dof_up_coop}
\begin{align}	
d^{\text{coop}}_{k}&\le\lim_{P\rightarrow \infty}\frac{\frac{1}{K-1}\sum_{n=1}^{N}\left[ \frac{\alpha }{2}\log\left( \frac{|r_{k}(n,n)|^2 \beta_{k,n} P}{\sigma_\text{R}^2}\right) \right]^+}{\log\big( \beta_{k,n} P\big)} \\
&=N\cdot\frac{\alpha}{2}\cdot\frac{1}{K-1}, \ \ k\in\mathcal{I}_K.
\end{align}
\end{subequations}	
Denote by $ \mathcal{D}^{\text{coop}}_{\text{up}} $ the uplink DoF region specified by \eqref{dof_up_coop}. From \eqref{equ:LP-Coop-c}, we have 	
\begin{subequations}\label{dof_down_coop}
\begin{align}	
\sum_{k^\prime\in \mathcal{I}_K\backslash k}d^{\text{coop}}_{k^\prime} &\le\lim_{P\rightarrow \infty}\frac{\frac{1-\alpha }{2}\log \left\vert \mathbf{I}_{M}+\frac{1}{\sigma_k^2}\mathbf{G}_{k}\mathbf{Q}_\text{R}\mathbf{G}
	_{k}^{\text{T}}\right\vert}{\log\big( \beta_\text{R} P\big)}\\
&=\lim_{P\rightarrow \infty}\frac{\frac{1-\alpha }{2}\log \left\vert \mathbf{I}_{M}+\beta_\text{R} P\frac{1}{\sigma_k^2}\mathbf{G}_{k}\mathbf{G}
	_{k}^{\text{T}}\right\vert}{\log\big( \beta_\text{R} P\big)}\label{35bb}\\
&=N\cdot\frac{1-\alpha}{2}, \ \ k\in\mathcal{I}_K,
\end{align}
\end{subequations}
where step \eqref{35bb} follows from the fact that the optimal $\mathbf{Q}_\text{R}$ at high SNR is given by
\begin{equation}\label{optimal_qr}
\mathbf{Q}_\text{R} = \beta_\text{R} P \mathbf{I}_N.
\end{equation}
Denote by $ \mathcal{D}^{\text{coop}}_{\text{down}} $ the downlink DoF region specified by \eqref{dof_down_coop}.

From \eqref{dof_up_dist} and \eqref{dof_down_dist1}, we readily obtain $ \mathcal{D}^{\text{dist}}_{\text{up}}\subset \mathcal{D}^{\text{dist}}_{\text{down}} $ for $ \alpha <\frac{1}{2} $. Similarly, from \eqref{dof_up_coop} and \eqref{dof_down_coop}, we obtain $ \mathcal{D}^{\text{coop}}_{\text{up}}\subset \mathcal{D}^{\text{coop}}_{\text{down}} $ for $ \alpha <\frac{1}{2} $. This means that the optimal sum-rates of both \eqref{equ:LP-Dist} and \eqref{equ:LP-Coop} are determined by the uplink at high SNR. Recall that the uplink constraints of the distributive relaying scheme are given by \eqref{equ:LP-Dist-b}, and those of the cooperative scheme are given by \eqref{equ:LP-Coop-b}. Clearly, \eqref{equ:LP-Dist-b} and \eqref{equ:LP-Coop-b} yield the same sum-rate. Therefore, the rate gap between the two schemes is zero at high SNR when $ \alpha <\frac{1}{2} $.

What remains is the case of  $ \alpha >\frac{1}{2} $. It can be seen that $ \mathcal{D}^{\text{dist}}_{\text{down}} $ is not necessarily contained in $ \mathcal{D}^{\text{dist}}_{\text{up}} $. However, by summing the inequalities in \eqref{dof_down_dist1} over  indexes  $ k $ and $ n $, we obtain
\begin{equation}\label{kkd}
\begin{split}	
\sum_{n=1}^{N}\sum_{k=1}^{K}d^{\text{dist}}_{k,n} \le N\cdot\frac{1-\alpha}{2}\cdot\frac{K}{K-1}, \ \ k\in\mathcal{I}_K, n\in \mathcal{I}_N.
\end{split}
\end{equation}
The equality of \eqref{kkd} gives the maximum downlink sum DoF, and can be achieved at $\{ d^{\text{dist}}_{k,n}=\frac{1-\alpha}{2}\cdot\frac{1}{K-1}|\  k\in\mathcal{I}_k,n\in\mathcal{I}_N \}$. It is also readily verified that the DoF tuple $ \{d^{\text{dist}}_{k,n}=\frac{1-\alpha}{2}\cdot\frac{1}{K-1}|\ k\in\mathcal{I}_k,n\in\mathcal{I}_N \} $ falls into the uplink DoF region  $ \mathcal{D}^{\text{dist}}_{\text{up}} $ when $ \alpha>\frac{1}{2} $. Therefore, the optimal sum DoF is determined solely by  $ \mathcal{D}^{\text{dist}}_{\text{down}} $. This further implies that in the high SNR regime, the optimal sum-rate of \eqref{equ:LP-Dist} is determined solely by the downlink constraints  \eqref{equ:LP-Dist-c}. Thus, the optimizing problem  \eqref{equ:LP-Dist} at high SNR for $ \alpha >\frac{1}{2} $ can be simplified as 
\begin{subequations} \label{redist}
	\begin{align}		
	\mathop{\text{maximize}}\limits_{\{R_{k,n}\} }\ \ \ \ &\sum_{k=\mathrm{1}}^{K} \sum_{n = 1}^N R_{k,n}\\
	\text{subject to}\ \ \ \
	&\sum_{n\in \mathcal{S}}\sum_{k^\prime\in \mathcal{I}_K\backslash k}R_{k^\prime,n} \leq \!\frac{1-\alpha }{2}\log\left\vert\mathbf{I}_M \!\!+\!\!\frac{1}{\sigma^2_k}\sum_{n\in \mathcal{S}}\beta_\text{R}P\mathbf{g}_{n,k}\mathbf{g}_{n,k}^{T} \right \vert,\ \ k \in \mathcal{I}_K, \mathcal{S}\subseteq \mathcal{I}_N.\label{redistc}
	\end{align}
\end{subequations}
Similarly, we see that the optimal sum-rate of \eqref{equ:LP-Coop} at high SNR is only determined by the downlink constraints \eqref{equ:LP-Coop-c} and  \eqref{equ:LP-Coop-d}, i.e., the optimization problem \eqref{equ:LP-Coop} can be rewritten as 
\begin{subequations} \label{recoop}
	\begin{align}		
	\mathop {\text{maximize}}\limits_{\mathbf{Q}_\text{R},\{R_k\} }\ \ \ \ &\sum_{k=\mathrm{1}}^{K}R_k \\
	\text{subject to}\ \ \ \
	&\sum_{k^\prime\in \mathcal{I}_K\backslash k}^{K}R_{k^\prime} \leq \frac{1-\alpha }{2}\log \left\vert \mathbf{I}_{M}+\frac{1}{\sigma_k^2}\mathbf{G}_{k}\mathbf{Q}_\text{R}\mathbf{G}
	_{k}^{\text{T}}\right\vert, \ k\in\mathcal{I}_K \label{recoopc} \\
	&\mathrm{tr} \{\mathbf{Q}_\text{R}\}\leq N\beta_\text{R}P,  \mathbf{Q}_\text{R}\succeq\mathbf{0}. \label{recoopd}
	\end{align}
\end{subequations}
From \eqref{optimal_qr}, we can further write \eqref{recoop} as 
\begin{subequations} \label{recoop-2}
	\begin{align}		
	\mathop {\text{maximize}}\limits_{\mathbf{Q}_\text{R},\{R_k\} }\ \ \ \ &\sum_{k=\mathrm{1}}^{K}R_k \\
	\text{subject to}\ \ \ \
	&\sum_{k^\prime\in \mathcal{I}_K\backslash k}^{K}R_{k^\prime} \leq \frac{1-\alpha }{2}\log\left\vert\mathbf{I}_M \!\!+\!\!\frac{1}{\sigma^2_k}\sum_{n=1}^{N}\beta_\text{R}P\mathbf{g}_{n,k}\mathbf{g}_{n,k}^{T} \right \vert,\ \ k \in \mathcal{I}_K. \label{recoopb-2}
	\end{align}
\end{subequations}

We now show that \eqref{redist} and \eqref{recoop-2} yields the same maximum sum-rate. This is true by noting that for any given $ k $, \eqref{redistc} gives the capacity region of a multiple access channel with $ K-1 $ users, with the optimal sum-rate given by the right hand side of \eqref{recoopb-2}. This concludes the proof.$\quad \blacksquare$

\emph{Remark 1:} Note that the case of $ \alpha=\frac{1}{2} $ is excluded in Theorem 2. In fact, when  $ \alpha=\frac{1}{2} $, there exists a non-vanishing gap between the distributive relaying scheme and the cooperative relaying scheme as the SNR goes to infinity. Nevertheless, this gap is usually very small. For example, in the settings of Fig. \ref{Figure alpha}, the rate gap at $ \alpha =\frac{1}{2} $ is only about 0.1 bit/channel use.

\emph{Remark 2:} In practical scenarios, relays are often scattered over a large area, and thus full cooperation among relays is difficult. Theorem 2 states that the performance penalty for distributive relaying is usually marginal as compared with cooperative relaying, especially in the high SNR region. This demonstrates the advantage of the proposed distributive relaying scheme.

\subsection{Asymptotic Optimality for $ K=2 $}
We show that the proposed lattice coding scheme is asymptotically optimal for $ K=2 $.  We have the following theorem.

\begin{theorem}\label{Theorem k=2}
When $ \alpha\in (0,\frac{1}{2})\cup(\frac{1}{2},1)$ and $ K=2 $, the proposed lattice coding scheme asymptotically achieves the sum-rate capacity as $ P/\sigma^2 \rightarrow \infty $. 
\end{theorem} 
\textit{Proof:}
	From Theorem 2, it suffices to show that the proposed scheme with cooperative relaying can achieve the cut-set bound as $ P/\sigma^2\rightarrow \infty $. From \eqref{bounddown} and \eqref{equ:LP-Coop-c}, we see that the proposed scheme achieves the cut-set outer bound in the downlink. Thus, it suffices to focus on the uplink. From \eqref{boundup} and $ K=2 $, we can express the uplink rate bound of user $ k $ at high SNR by
	\begin{subequations}
		\begin{eqnarray}
		R_{k} &\le& \max_{\mathrm{tr}\{\mathbf{Q}_{k}\}\leq \beta_k P}\frac{\alpha}{2}\log \left\vert \mathbf{I}_{N}+\frac{1}{\sigma^2}\mathbf{H}_{k}\mathbf{Q}_{k}\mathbf{H}_{k}^{\text{T}}\right\vert\label{38aa}\\
		&=& \frac{\alpha}{2}\log \left\vert \frac{\beta_k P}{N\sigma^2}\mathbf{H}_{k}\mathbf{H}_{k}^{\text{T}}\right\vert\, \  \mathrm{for}\ k = 1, 2,
		\end{eqnarray}
	\end{subequations}
	where \eqref{38aa} follows by substituting $ P_k=\beta_k P $ and $ \sigma^2_{\text{R}}=\sigma^2 $ into \eqref{boundup}.
	At high SNR, the achievable rate of user $ k $ of the proposed scheme is given by \eqref{equ:LP-Coop-b}:
	\begin{subequations} \label{ikk}
	\begin{align}	
	R_{k}&\le \sum_{n=1}^{N}\frac{\alpha}{2}\log\left( \frac{|r_{k}(n,n)|^2 \beta_k P }{N\sigma^2}\right)\label{ikka}\\
	&= \frac{\alpha}{2}\log \left| \frac{\beta_k P}{N\sigma^2}\mathbf{R}_{k}\mathbf{R}_{k}^{\text{T}}\right| \label{ikkb}\\
	&= \frac{\alpha}{2}\log \left| \frac{\beta_k P}{N\sigma^2}\mathbf{H}_{k}\mathbf{H}_{k}^{\text{T}}\right|\ \  \mathrm{for}\ k = 1, 2,\label{ikkc}
	\end{align}
	\end{subequations}
	where \eqref{ikka} assumes equal power allocation, \eqref{ikkb} utilizes the fact that $ \mathbf{R}_k $ is upper-triangular and obtained from the RQ decomposition shown in  \eqref{hhrr}.  Therefore, the proposed scheme achieves the cut-set bound in the high SNR regime, which concludes the proof.$\qquad \qquad \qquad \qquad\quad  \blacksquare $

With $ K=2 $, multiway relaying reduces to two-way relaying. Previously, GSVD-based lattice-coding schemes were proposed in \cite{YangIT11} to approach the asymptotic capacity of the MIMO two-way relay channel. However, GSVD requires relay cooperations and therefore cannot be applied to distributive relaying. To the best of our knowledge, Theorem \ref{Theorem k=2} is the first to achieve the asymptotic capacity of the MIMO two-way relay channel with distributed relays.

\subsection{ Asymptotic Sum-Rate Gap for $ K\ge 3 $ }
We now consider the case of $ K\geq 3 $. In this case, our proposed scheme cannot achieve the sum-capacity upper bound in general. Our goal is to analyze the average sum-rate gap between the proposed scheme and the sum capacity upper bound. 

To start with, we assume the elements of the channel matrices $ \{\mathbf{H}_k\} $ and $ \{\mathbf{G}_k\} $ are independently and identically drawn from $ \mathcal{N}(0,1) $. Let $ \Delta $ be the average sum-rate gap between the maximum sum-rate of the distributive relaying scheme given by \eqref{equ:LP-Dist} and the cut-set bound in \eqref{bound} as $ P/\sigma^2\to \infty $. Denote 
\begin{equation}\label{dn}
d_n=(K-1)M-N+n.
\end{equation}
Then, we have the following result.
\begin{theorem} \label{Theorem gap}
	Consider the MIMO MDRC with $ K\geq 3 $ and $ \beta_1=\beta_2=\cdots\beta_K=\beta $. For $ \alpha > \frac{1}{2} $, $ \Delta= 0. $ For $ \alpha < \frac{1}{2} $, the asymptotic rate gap is bounded by
\begin{equation}\label{gap-o}
	\begin{split}
		\Delta\le \frac{\alpha NK(K-2)}{2(K-1)^2}+\frac{K}{K-1}\sum_{n=1}^{N}\bigg(\alpha\cdot\log e\cdot \mathrm{E}\Big[\mathrm{FisherZ}(d_n,n)\Big]+\frac{\alpha}{2}\log\frac{d_n}{n}\bigg),
	\end{split}	
	\end{equation} 	
	where $ \mathrm{FisherZ}(d_n,n) $ represents the Fisher's Z-distribution with degree $ d_n $ and $ n $.
\end{theorem}
	\textit{Proof:} See Appendix \ref{sec appendixb}.

\emph{Remark 3:} From \eqref{gap-o}, we see that the sum-rate gap does not scale with the transmission power $ P $ in the high SNR region. This implies that our proposed scheme statistically achieves the sum capacity of the MIMO MDRC within a constant gap in the high SNR regime.

\emph{Remark 4:} When  both $ M$ and $N $ are large, the  Fisher's Z-distribution in \eqref{gap-o} can be approximated by a normal distribution with mean $ \frac{1}{n}-\frac{1}{d_n} $ \cite{Fisher}. Then,  \eqref{gap-o} becomes
\begin{equation}
	\begin{split}
	\Delta\le \frac{\alpha NK(K-2)}{2(K-1)^2}+\frac{K\alpha}{K-1}\log e\sum_{n=1}^{N}(\frac{1}{n}-\frac{1}{d_n})+\frac{K\alpha}{2(K-1)}\sum_{n=1}^{N}\log\frac{d_n}{n}.\label{equ:gap}
	\end{split}	
\end{equation} 	
 As $K$ increases, the gap scales in the order of $\log K$.

\section{Numerical Results}\label{sec num}
\begin{figure}[!t]
	\centering
	\includegraphics[width = 0.6\textwidth]{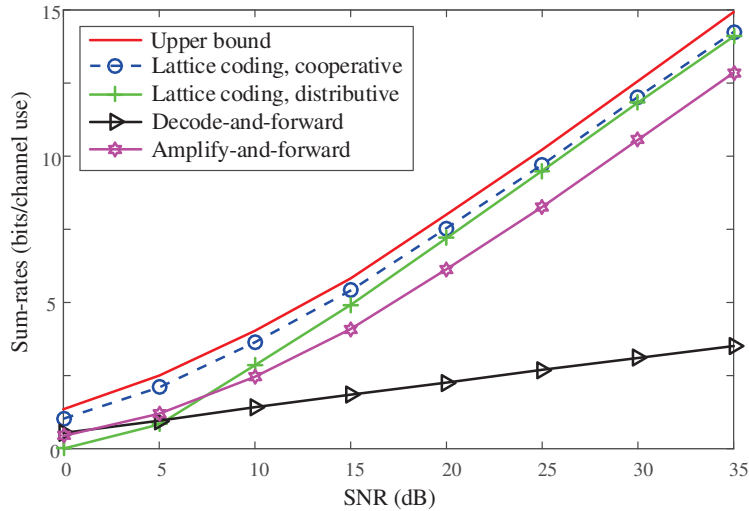}
	\caption{Performance comparison with power optimization among different coding schemes with
		$M = N = 4, K = 3$, $ \alpha=\frac{1}{2} $,  $P_k = P$ and $P_{\text{R},n} = P/N$.}
	\label{Figure all}
\end{figure}

\begin{figure}[!t]
	\centering
	\includegraphics[width = 0.6\textwidth]{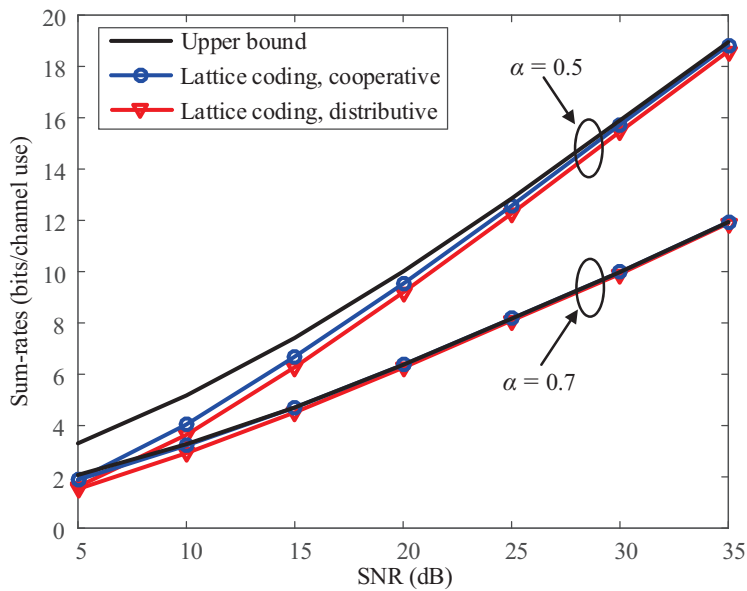}
	\caption{Performance comparison between the cut-set bound and lattice coding scheme with distributive and cooperative relays with $M=N=4$, $K=2 $, $P_k = P$ and $P_{\text{R},n} = P/N$. }
	\label{Figure kk}
\end{figure}

\begin{figure}[!t]
	\centering
	\includegraphics[width = 0.6\textwidth]{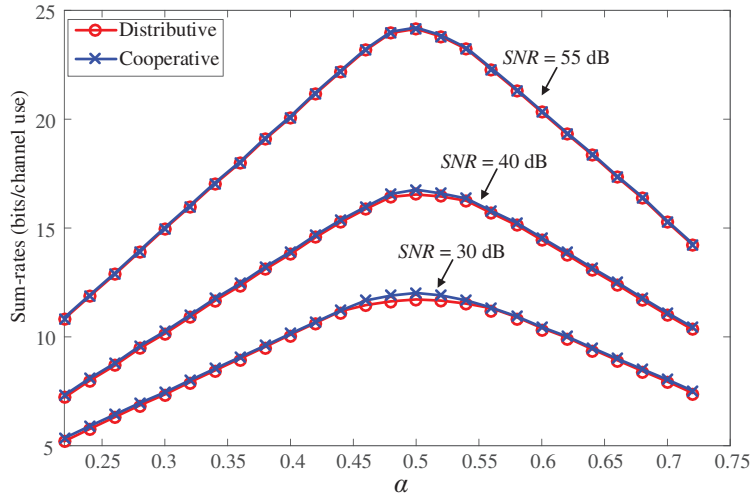}
	\caption{Performance comparison between the distributive case and the cooperative case with $M=N=4$, $ K=3 $, $P_k = P$ and $P_{\text{R},n} = P/N$. }
	\label{Figure alpha}
\end{figure}

We now present the  performance comparisons of various relaying schemes. In simulation, we set $ M = N = 4, K = 3, \alpha =\frac{1}{2} $, $\sigma_{\text{R}}^2 =\sigma_k^2 = 1 $, and $P_{\text{R},n} = P_k = P, \text{for}\ n\in \mathcal{I}_N, k\in \mathcal{I}_K $. The SNR is defined as $ P/\sigma^2 $. From Fig. \ref{Figure all}, the proposed lattice coding scheme with distributive relays performs within a marginal gap to the case with cooperative relays in the high SNR region (though Theorem 2 does not hold for $ \alpha=\frac{1}{2} $). The proposed distributed relaying scheme performs within a gap of about 0.5 dB to the cut-set upper bound. We also see that the lattice coding scheme considerably outperforms the reference schemes such as decode-and-forward (DF) and amplify-and-forward (AF) which are detailed in Appendix \ref{sec appendixa}. 

Fig. \ref{Figure kk} shows the achievable sum-rate of the proposed nested lattice coding schemes with distributive relays and with cooperative relays. The simulation settings are as following: $M=N=4$, $K=2 $, $P_k = P$ and $P_{\text{R},n} = P/N$, $ \alpha=\frac{1}{2} \text{ or }\frac{7}{10} $. The cut-set bound in \eqref{bound} is also included for comparison. From Fig. \ref{Figure kk}, when $ K=2 $ and $ \alpha =\frac{7}{10} $, the distributive relaying asymptotically achieves the same performance as cooperative relaying at high SNR regime, which verifies Theorem \ref{theorem_DC}. Furthermore, distributive relaying asymptotically achieves the cut-set bound, which indicates that the proposed scheme is asymptotically optimal when $ K=2$ and $\alpha =\frac{7}{10} $, which agrees with Theorem \ref{Theorem k=2}. We also see that for $ \alpha =\frac{1}{2} $, the proposed scheme cannot achieve the cut-set bound. But the performance gap is marginal in the high SNR regime.

Fig. \ref{Figure alpha} compares the sum-rates of the distributive relaying scheme and the cooperative relaying scheme against $ \alpha $, with the simulation settings of $M=N=4$, $ K=3 $, $P_k = P$ and $P_{\text{R},n} = P/N$. We see that the maximum sum-rates of both schemes are achieved at $ \alpha=\frac{1}{2} $. We also see that as the SNR increases, the rate gap between the two schemes vanishes for $ \alpha\neq\frac{1}{2} $, which agrees with Theorem 2. For $ \alpha=\frac{1}{2} $, the rate gap is stuck at around 0.1 bit per channel use as the SNR tends to infinity.

\section{Conclusion}\label{sec con}

In this paper, we considered efficient system design for the MIMO MDRC with full data exchange. We proposed a  nested lattice coding scheme and derived an achievable rate region for the proposed scheme. We showed that distributive relaying incurs no sum-rate loss at high SNR for the time splitting factor $ \alpha \neq \frac{1}{2} $, as compared with cooperative relaying. We also showed that the proposed scheme achieves the sum capacity at high SNR within a constant gap. Numerical results were also presented to verify the theoretical analysis and show the performance advantage of the proposed scheme compared to other existing schemes. How to narrow the gap between our scheme and the cut-set bound will be an interesting topic for future research.

\appendices
\section{Proof of Theorem 4}\label{sec appendixb}
From Theorem 2, the proposed distributive relaying scheme can achieve the same sum-rate as the cooperative relaying scheme when $ \alpha \neq \frac{1}{2} $. Thus, to prove Theorem 4, it suffices to consider the rate gap between the cooperative relaying scheme with the optimal sum-rate given by \eqref{equ:LP-Coop} and the upper bound in \eqref{bound}.

Similarly to the proof of Theorem 2, we start with analyzing the DoF region of the upper bound in \eqref{bound}. Denote by $ d^{\text{bound}}_{k} $ the DoF of user $ k $ based on the upper bound in \eqref{bound}. Then for $ k\in\mathcal{I}_K $,
\begin{subequations}\label{dof_up_bound}
\begin{align}	
\sum_{k^\prime\in \mathcal{I}_K\backslash k}d^{\text{bound}}_{k}&\le \lim_{P\rightarrow \infty} \frac{\frac{\alpha}{2}\log\Big|\mathbf{I}_N+\frac{1}{\sigma_k^2}\sum_{k^\prime\neq k}\mathbf{H}_{k^\prime}\mathbf{Q}_{k^\prime}\mathbf{H}_{k^\prime}^{\text{T}}\Big|}{\log (\beta_k P)}\\
&=N\cdot\frac{\alpha}{2},
\end{align}
\end{subequations}
and
\begin{subequations}\label{dof_down_bound}
\begin{align}	
\sum_{k^\prime\in \mathcal{I}_K\backslash k}d^{\text{bound}}_{k^\prime} &\le\lim_{P\rightarrow \infty}\frac{\frac{1-\alpha }{2}\log \left\vert \mathbf{I}_{M}+\frac{1}{\sigma_k^2}\mathbf{G}_{k}\mathbf{Q}_\text{R}\mathbf{G}
	_{k}^{\text{T}}\right\vert}{\log\big( \beta_\text{R} P\big)}\\
&=N\cdot\frac{1-\alpha}{2}.
\end{align}
\end{subequations}
Denote by $ \mathcal{D}^{\text{bound}}_{\text{up}} $ the uplink DoF region specified by \eqref{dof_up_bound} and by $ \mathcal{D}^{\text{bound}}_{\text{down}} $ the downlink DoF region specified by \eqref{dof_down_bound}. Also denote the expected values of the right hand side of \eqref{bound} and \eqref{crate} as  
\begin{subequations}\label{ii}
	\begin{align}
	R^{\text{bound}}_{\text{up}}&\triangleq \mathrm{E}\bigg[\max_{\mathrm{tr}\{\mathbf{Q}_{k^\prime}\}\le P_k}\frac{\alpha}{2}\log\Big|\mathbf{I}_N+\frac{1}{\sigma^2}\sum_{k^\prime\neq k}\mathbf{H}_{k^\prime}\mathbf{Q}_{k^\prime}\mathbf{H}_{k^\prime}^{\text{T}}\Big|\bigg]\label{iia}\\
	R^{\text{bound}}_{\text{down}}&\triangleq \mathrm{E}\bigg[\max_{\mathrm{tr}\{\mathbf{Q}_\text{R}\}\le P_\text{R}}\frac{1-\alpha}{2}\log\Big|\mathbf{I}_M+\frac{1}{\sigma^2}\mathbf{G}_{k}\mathbf{Q}_\text{R}\mathbf{G}_{k}^{\text{T}}\Big|\bigg]\label{iib}\\
	R^{\text{coop}}_{\text{up},k}&\triangleq\mathrm{E}\bigg[\max_{\{P_{k^\prime,n}\}}\sum_{\substack{k^\prime \in \mathcal{I}_K\backslash k}}\frac{1}{K-1}\sum_{n=1}^{N}\left[ \frac{\alpha}{2}\log\left( \frac{|r_{k}(n,n)|^2 P_{k,n} }{\sigma^2}\right) \right]^+\bigg],\ k \in \mathcal{I}_K\label{iic}\\
	R^{\text{coop}}_{\text{down}}&\triangleq\mathrm{E}\bigg[\max_{\mathrm{tr}\{\mathbf{Q}_\text{R}\}\le P_\text{R}}\frac{1-\alpha}{2}\log \left\vert \mathbf{I}_{M}+\frac{1}{\sigma^2}\mathbf{G}_{k}\mathbf{Q}_\text{R}\mathbf{G}
	_{k}^{\text{T}}\right\vert\bigg],\label{iid}
	\end{align}
\end{subequations}
where the expectation is taken over the channel matrices. Note that $ R^{\text{bound}}_{\text{up}}, R^{\text{bound}}_{\text{down}},$ and $ R^{\text{coop}}_{\text{down}}$ are invariant to the user index $ k $ due to user symmetry. Also note that $ R^{\text{bound}}_{\text{up}}, R^{\text{bound}}_{\text{down}}, $ and $R^{\text{coop}}_{\text{down}}$ are optimal sum-rates of $ K-1 $ users. Thus, we need to include a multiplicative factor of $ \frac{K}{K-1} $ to express the sum-rates of $ K $ users. For example, $ \frac{K}{K-1}R^{\text{bound}}_{\text{up}} $ represents the average uplink sum-rate of the upper bound.

When $ \alpha>\frac{1}{2} $, we have $ \mathcal{D}^{\text{bound}}_{\text{down}} \subset \mathcal{D}^{\text{bound}}_{\text{up}}  $. This implies that the optimal sum-rate of the upper-bound is  determined by the downlink phase. Recall from the proof of Theorem 2 that the optimal sum-rate of the cooperative relaying scheme is also determined by the downlink phase. Thus, using \eqref{iib} and \eqref{iid}, we have
\begin{equation}
	\begin{split}
		\Delta=\frac{K}{K-1}\Big(R^{\text{bound}}_{\text{down}}-R^{\text{coop}}_{\text{down}}\Big)=0.
	\end{split}
\end{equation} 

We next focus on the case of $ \alpha <\frac{1}{2} $. From \eqref{dof_up_bound} and \eqref{dof_down_bound}, we have  $ \mathcal{D}^{\text{up}}_{\text{bound}}\subset \mathcal{D}^{\text{down}}_{\text{bound}} $ for  $ \alpha <\frac{1}{2} $. Together with the proof of Theorem 2, we see that the sum-rates of the upper-bound and the cooperative relaying scheme are both determined by the uplink phase. Thus, $\Delta=\frac{K}{K-1} R^{\text{bound}}_{\text{up}}-\frac{1}{K-1}\sum_{k=1}^{K}R^{\text{coop}}_{\text{up},k}$. Denote
	\begin{align}
		\mathbf{\widetilde{H}}_{k}&=[\mathbf{H}_1,\cdots,\mathbf{H}_{k-1},\mathbf{H}_{k+1},\cdots,\mathbf{H}_K]\\
		\mathbf{\widetilde{Q}}_{k}&=\mathrm{blkdiag}\{\mathbf{Q}_{1},\mathbf{Q}_{2},\cdots,\mathbf{Q}_{k-1},\mathbf{Q}_{k+1},\cdots,\mathbf{Q}_{K}\} ,
	\end{align}
where $ \mathrm{blkdiag}\{\cdot\} $ stands for a block-diagonal matrix formed by the given blocks. Then
\begin{subequations}\label{aty}
	\begin{align}
	R^{\text{bound}}_{\text{up}}&=\mathrm{E}\Bigg[\max_{\substack{\mathbf{\widetilde{Q}}_{k} \text{ is blockdiagonal}\\\text{tr}\{\mathbf{Q}_k\}\le \beta P, \forall k}}\frac{\alpha}{2}\log\bigg|\mathbf{I}_N+\frac{1}{\sigma^2}\mathbf{\widetilde{H}}_{k}\mathbf{\widetilde{Q}}_{k}\mathbf{\widetilde{H}}_{k}^{\text{T}}\bigg|\Bigg]\label{aty-a}\\
	&\le\mathrm{E}\Bigg[\max_{\text{tr}\{\mathbf{Q}_k\}\le \beta P, \forall k}\frac{\alpha}{2}\log\bigg|\mathbf{I}_N+\frac{1}{\sigma^2}\mathbf{\widetilde{H}}_{k}\mathbf{\widetilde{Q}}_{k}\mathbf{\widetilde{H}}_{k}^{\text{T}}\bigg|\Bigg]\label{aty-b}\\
	&\le\mathrm{E}\Bigg[\max_{\text{tr}\{\mathbf{\widetilde{Q}}_{k}\}\le (K-1)\beta P}\frac{\alpha}{2}\log\bigg|\mathbf{I}_N+\frac{1}{\sigma^2}\mathbf{\widetilde{H}}_{k}\mathbf{\widetilde{Q}}_{k}\mathbf{\widetilde{H}}_{k}^{\text{T}}\bigg|\Bigg],\label{aty-c}
	\end{align}
\end{subequations}
where \eqref{aty-b} follows by relaxing $ \mathbf{\widetilde{Q}}_{k} $ to a full matrix, and \eqref{aty-c} follows by relaxing the per-user power constraint to a total power constraint.

The singular value decomposition (SVD) of $ \mathbf{\widetilde{H}}_{k} $ is given by $ \mathbf{\widetilde{H}}_{k}=\mathbf{\widetilde{U}}_{k}\mathbf{\widetilde{D}}_{k}\mathbf{\widetilde{V}}_{k}^{\text{T}} $. Then, in the high SNR regime, the asymptotically optimal $ \mathbf{\widetilde{Q}}_{k} $ to (\ref{aty}) is given by
\begin{equation}\label{ggd}
\mathbf{\widetilde{Q}}_{k}=\frac{(K-1)\beta P}{N}\mathbf{\widetilde{V}}_{k}\left[
\begin{matrix}
\mathbf{I}_N &\mathbf{0}\\
\mathbf{0}&\mathbf{0}
\end{matrix}
\right]\mathbf{\widetilde{V}}^{\text{T}}_{k}.
\end{equation}
Substituting (\ref{ggd}) into (\ref{aty}), we obtain
\begin{subequations}\label{i6}
\begin{align}
R^{\text{bound}}_{\text{up}}&\leq \mathrm{E}\Bigg[\frac{\alpha}{2}\log\bigg|\mathbf{I}_N+\frac{1}{\sigma^2}\frac{(K-1)\beta P}{N}\mathbf{\widetilde{H}}_{k}\mathbf{\widetilde{H}}_{k}^{\text{T}}\bigg|\Bigg]\\
&=\frac{N\alpha}{2}\log\frac{(K-1)\beta P}{N\sigma^2}+\frac{\alpha}{2}\mathrm{E}\Big[\log\big|\mathbf{\widetilde{H}}_{k}\mathbf{\widetilde{H}}_{k}^{\text{T}}\big|\Big]+o(1),
\end{align}
\end{subequations}
where $ o(1) \to 0 $ as $ P\to \infty $. We now present a lower bound of $ R^{\text{coop}}_{\text{up},k} $ by assuming  $ P_{k,n}=\frac{K-1}{2N} \beta P \text{ for } k=2,\cdots,K$ and $ P_{k,n}=\frac{K-1}{N} \beta P \text{ for } k=1 \text{ or } K $. Specifically, we have
\begin{align}
R^{\text{coop}}_{\text{up},k} &\ge\mathrm{E}\bigg[\sum_{\substack{k^\prime\neq k\\k^\prime=\{1,K\}}}\frac{\alpha}{2(K-1)}\sum_{n=1}^{N}\log\Big(\frac{K-1}{N}\cdot\frac{\beta P}{\sigma^2}|r_{k^\prime}(n,n)|^2\Big)\nonumber\\
&\qquad +\sum_{\substack{k^\prime\neq k\\k^\prime\neq\{1,K\}}}\frac{\alpha}{2(K-1)}\sum_{n=1}^{N}\log\Big(\frac{K-1}{2N}\cdot\frac{\beta P}{\sigma^2}|r_{k^\prime}(n,n)|^2\Big)\bigg]\nonumber\\
&=\mathrm{E}\bigg[\sum_{k^\prime\in \mathcal{I}_K\backslash k}^{K}\frac{\alpha}{2(K-1)}\sum_{n=1}^{N}\log\Big(\frac{K-1}{N}\cdot\frac{\beta P}{\sigma^2}|r_{k^\prime}(n,n)|^2\Big)
\bigg]-\frac{\gamma_k N\alpha}{2(K-1)}\nonumber\\
&\ge\frac{N\alpha}{2}\log\Big(\frac{(K-1)\beta P}{N\sigma^2}\Big)\!+\!\mathrm{E}\bigg[\sum_{k^\prime\in \mathcal{I}_K\backslash k}^{K} \frac{\alpha}{2(K-1)}\sum_{n=1}^{N}\log |r_{k^\prime}(n,n)|^2\bigg]\!-\!\frac{\alpha N(K\!-\!2)}{2(K\!-\!1)}\nonumber\\
&\triangleq R^{\text{coop}}_{\text{up}},\label{i7}
\end{align}
where $ \gamma_k=K-2 $ for $ k=1 \text{ or } K $, and  $ \gamma_k=K-3 $ otherwise. Note that $ R^{\text{coop}}_{\text{up}} $ in \eqref{i7} is invariant to the index $ k $. Let the RQ decomposition of $ \widetilde{\mathbf{H}}_{k} $ be $ \widetilde{\mathbf{H}}_{k}=\widetilde{\mathbf{R}}_{k}\widetilde{\mathbf{Q}}_{k} $. From (\ref{i6}) and (\ref{i7}),
\begin{align}
\frac{K-1}{K}\Delta &\le R^{\text{bound}}_{\text{up}}-R^{\text{coop}}_{\text{up}}\nonumber \\
&\le\frac{\alpha N(K-2)}{2(K-1)}+\frac{\alpha}{2(K-1)}\mathrm{E}\Big[\sum_{k^\prime\in \mathcal{I}_K\backslash k}^{K}\Big(\log\big|\mathbf{\widetilde{H}}_{k}\mathbf{\widetilde{H}}_{k}^{\text{T}}\big|-\sum_{n=1}^{N}\log|r_{k^\prime(n,n)}|^2\Big)\Big]+o(1)\nonumber\\
&=\frac{\alpha N(K-2)}{2(K-1)}+\frac{\alpha}{2(K-1)}\mathrm{E}\bigg[\sum_{k^\prime\in \mathcal{I}_K\backslash k}^{K}\sum_{n=1}^{N}\log\frac{|t_{k^\prime}(n,n)|^2}{|r_{k^\prime}(n,n)|^2}\bigg],\label{i9}
\end{align}
where  $ t_k(n,n) $ represent the $ n $-th diagonal element of $ \mathbf{\widetilde{R}}_k $. Recall that the elements of $ \mathbf{\widetilde{H}}_k $, $ k=1,\cdots,K $ are independently drawn from a common Gaussian distribution. Thus, $ |r_{k}(n,n)|^2 $  follows the $ \chi $-square distribution with degree $ d_n $ in \eqref{dn},
and $ t_{k^\prime}(n,n) $ follows the $ \chi $-square distribution with degree $ n $. Thus, 
\begin{equation*}
\frac{1}{2}\ln\bigg(\frac{|t_{k}(n,n)|^2/n}{|r_{k}(n,n)|^2/d_n}\bigg)\sim\text{FisherZ}(d_n,n),
\end{equation*}
where $ \text{FisherZ}(d_n,n) $ represents the Fisher's Z-distribution with degrees $ d_n $ and $ n $ \cite{Fisher}. Then, from (\ref{i9}), we have
\begin{subequations}
\begin{align}\label{i10}\nonumber
\frac{K-1}{K}\Delta &\le\frac{\alpha N(K-2)}{2(K-1)}+\frac{\alpha}{2(K-1)}\mathrm{E}\bigg[\sum_{k^\prime\in \mathcal{I}_K\backslash k}^{K}\sum_{n=1}^{N}\bigg(\log\frac{|r_{k^\prime}(n,n)|^2/d_n}{|t_{k^\prime}(n,n)|^2/n}+\log \frac{d_n}{n}\bigg)\bigg]\\
&=\frac{\alpha N(K-2)}{2(K-1)}+\sum_{n=1}^{N}\bigg(\alpha\log e\  \mathrm{E}\Big[\text{FisherZ}(d_n,n)\Big]+\frac{\alpha}{2}\log\frac{d_n}{n}\bigg).\nonumber
\end{align}
\end{subequations}
Therefore, we obtain \eqref{gap-o}, which concludes the proof.$ \qquad \qquad \qquad \qquad \qquad \qquad\qquad \qquad\blacksquare $

\section{Sum-Rate Maximization for the AF and DF Strategies}\label{sec appendixa}
For comparison, we describe two alternative schemes based on the decode-and-forward (DF) and amplify-and-forward (AF) strategies. In the DF scheme, each relay $ n $ completely decodes all the messages from the $K$ users based on the received signal in (\ref{eq.yRn}). The corresponding achievable rate tuple of the uplink is constrained by the capacity region of the $K$-user MIMO multiple access channel specified in \eqref{eq.yRn}. The sum-rate maximization problem for the DF scheme can be formulated as

\begin{subequations}\label{dfrate}
	\begin{align} 
	\mathop {\text{maximize}}\limits_{\{\mathbf{Q}_k\}}\ \ \  &\sum_{k=\mathrm{1}}^{K}R_{k,n}\\
	\text{subject to}\ \ 
	&\mathrm{tr}\{\mathbf{Q}_k\}\leq P_k, \mathbf{Q}_k\succeq\mathbf{0}, \ k \in \mathcal{I}_K\\
	&\sum_{k\in \mathcal{S}}\! R_{k}\!\le\!
	\frac{\alpha }{2}\!\log\left|1\!+\!\frac{1}{\sigma_{\text{R}}^2}\!\sum_{k\in \mathcal{S}}\mathbf{h}_{k,n}\mathbf{Q}_k\mathbf{h}^{\text{T}}_{k,n}\right|
	,\ n\!\in\! \mathcal{I}_N,  \mathcal{S}\!\subseteq\! \mathcal{I}_K\ \\
	&\sum_{k^\prime \in\mathcal{I}_K\backslash k}^{K}\!\! R_{k^\prime} \leq \frac{1-\alpha }{2}\log \left\vert \mathbf{I}_{M}+\frac{1}{\sigma_k^2}\mathbf{G}_k\mathbf{Q}_\text{R}\mathbf{G}^{\text{T}}_k\right\vert,\  \ k\in\mathcal{I}_K.\label{dfrate-d}
	\end{align}
\end{subequations}
Note that the downlink rate constraint \eqref{dfrate-d} is identical to the downlink constraint in 
\eqref{equ:LP-Coop-c}. The reason is that with DF, each relay knows all the messages from the $ K $ users, and therefore full relay cooperation can be realized.

In the AF scheme, the operation of each relay $ n $ is to multiply its received signal by a scaling factor $ a_n, n\in\mathcal{I}_N $, so as to meet the transmission power budget $P_{\text{R},n}$. The corresponding achievable sum-rate is given by 
\begin{equation}\label{afrate_long}
\sum_{k\in \mathcal{S}} R_k\le\frac{1}{2}\log\frac{\left|\sum_{k\in \mathcal{S}}  \mathbf{H}_{k'}\mathbf{A}\mathbf{H}_{k}\mathbf{Q}_k \mathbf{H}^{\text{T}}_{k}\mathbf{A}^{\text{T}}\mathbf{H}_{k'}^{\text{T}}+\sigma_R^2\mathbf{G}_{k'}\mathbf{A}\mathbf{A}^{\text{T}}\mathbf{G}^{\text{T}}_{k'}+\sigma^2_{k'}\mathbf{I}\right|}{\left|\sigma_R^2\mathbf{G}_{k'}\mathbf{A}\mathbf{A}^{\text{T}}\mathbf{G}^{\text{T}}_{k'}+\sigma^2_{k'}\mathbf{I}\right|},\ \forall k'\in \mathcal{I}_K, \mathcal{S}\subseteq \mathcal{I}_K\backslash k',\\			
\end{equation}
where $ \mathbf{A}=\text{diag}\{a_1,a_2,\cdots,a_N\} $. Then the sum-rate maximization problem can be formulated as


\begin{subequations}\label{afrate}
	\begin{align} 	
	\mathop {\text{maximize}}\limits_{\{\mathbf{Q}_k\},\{a_n\}}\ \ \ \  &\sum_{k=\mathrm{1}}^{K}R_k\qquad \qquad \qquad \qquad \qquad \qquad \ \ \ \ \ \ \  \\
	\text{subject to}\ \ \ \ 	&(\ref{afrate_long}), \mathrm{tr}\{\mathbf{Q}_k\}\leq P_k,\ \mathbf{Q}_k\succeq\mathbf{0} \\
	&|a_n|^2\Big(\sum_{k=1}^{K}\mathbf{h}_{k,n}\mathbf{Q}_k\mathbf{h}^{\text{T}}_{k,n}\!+\!\sigma_{\text{R}}^{\mathrm{2}}\Big)\!\!\leq\! P_{\text{R},n},\ n \in \mathcal{I}_N.\ \ 
	\end{align}
\end{subequations}
Note that both (\ref{dfrate}) and (\ref{afrate}) can be solved by using standard convex programming.


\ifCLASSOPTIONcaptionsoff
  \newpage
\fi

%
%
%
%
\end{document}